%% file: p.tex
\documentclass[acmsmall]{acmart}
\usepackage{epsfig,endnotes}
\usepackage{subfiles}
\usepackage{subcaption}
\usepackage{multirow}
\usepackage{xcolor}
\usepackage{enumitem}

\makeatletter
\renewcommand*{\@fnsymbol}[1]{%
  \ensuremath{%
    \ifcase#1\or
      \dagger\or
      \ddagger\or
      \mathsection\or
      \mathparagraph\or
      \|\or
      **\or
      \dagger\dagger\or
      \ddagger\ddagger
    \else
      \@ctrerr
    \fi
  }%
}
\makeatother

\input{preamble}

\begin{document}

\date{}

\settopmatter{printfolios=true}

\title{Detecting and Characterizing \MassivelyShared IP Addresses}

 \author{Amanda Hsu }
 \email{ahsu67@gatech.edu}
 \affiliation{
 	\institution{Georgia Institute of Technology, Akamai}
 	\country{USA}
 }
  \author{Paul Pearce}
 \email{pearce@gatech.edu}
 \affiliation{
 	\institution{University of California, Irvine}
 	\country{USA}
 }
 \author{Frank Li}
 \email{frankli@gatech.edu}
 \affiliation{
 	\institution{Georgia Institute of Technology}
 	\country{USA}
 }

 \author{Arthur Berger}
 \authornote{Our colleague, mentor, and friend, Arthur, passed away on July 12, 2026. He reviewed the content of this paper prior to his passing.}
 \email{awberger@akamai.com}
 \affiliation{
 	\institution{Akamai, Massachusetts Institute of Technology}
 	\country{USA}
 }
 \author{Philipp Richter}
 \email{prichter@akamai.com}
 \affiliation{
 	\institution{Akamai}
 	\country{USA}
 }

\input{abstract}

\maketitle

\input{introduction}

\input{related_work}

\input{data}
\input{method}
\input{results}

\input{cloud_results}

\input{results_over_time}
\input{conclusion}

\input{acknowledgments}

\bibliographystyle{ACM-Reference-Format}
\bibliography{refs}

\appendix
\input{appendix}

\end{document}

%% file: preamble.tex
\newcommand{\ignore}[1]{}

\usepackage{xspace}
\usepackage{xcolor}
\usepackage{graphicx}
\usepackage{makecell}
\usepackage{wrapfig}
\usepackage[most]{tcolorbox}

\tcbset{
	mytakeaway/.style={
		colback=gray!20,     
		colframe=black!50,   
		arc=1mm,     
		boxrule=0.5pt,       
		width=\linewidth,    
		left=1mm,            
		right=1mm,           
		top=1mm,             
		bottom=1mm          
	}
}

\newcommand*{\eg}{e.g.,\@\xspace}
\newcommand*{\ie}{i.e.,\@\xspace}

\newcommand{\massivelyshared}{massively shared\@\xspace}
\newcommand{\Massivelyshared}{Massively shared\@\xspace}
\newcommand{\MassivelyShared}{Massively Shared\@\xspace}
\newcommand{\massivesharing}{massive sharing\@\xspace}
\newcommand{\MassiveSharing}{Massive Sharing\@\xspace}

\newcommand{\massive}{massive\@\xspace}
\newcommand{\nonmassivelyshared}{non-massively shared\@\xspace}
\newcommand{\PP}[1]{%
  \vspace{2px}%
  \noindent{\bf \begingroup\noexpandarg\IfEndWith{#1}{.}{#1}{#1.}\endgroup}%
}

\def\Snospace~{\S{}}

\makeatletter

%% file: abstract.tex
\begin{abstract}
\normalsize
IP addresses are commonly shared across devices and users for a variety of reasons, including NAT and proxies. 
These technologies operate at different scales, from residential NATs that share an IP address across devices in a home to large-scale Carrier Grade NATs that share hundreds or thousands of users on a single IP. 
Cases of large-scale IP sharing are distinct as they have significant implications for IP-based mechanisms 
such as attribution, blocklisting, and rate-limiting, where the consequences of mishandling affect a large quantity of end-users and organizations. 

In this work, we detect and characterize IP addresses shared at large scales, which we coin \massivelyshared. 
Leveraging diurnal patterns in traffic shape, we use data from a large CDN to characterize these IPs globally. 
We broadly find that \massive IP sharing is responsible for a large fraction of IPv4 traffic, concentrated in a small fraction of address space, with over 40\% of total traffic coming from less than 2\% of active IP addresses. 
We observe distinct patterns in deployment geographically, with particularly high rates of \massivelyshared traffic from some smaller countries. 
Comparatively, in IPv6, we find far fewer \massivelyshared addresses with some surprising exceptions among mobile providers. 
We additionally contextualize these addresses by other network characteristics, including identifying cellular connectivity and dual-stack capabilities, and identifying several instances of \massivelyshared IPs in proxy services hosted on cloud networks. 
Finally, we find that rates of \massivelyshared traffic are increasing over time, predicting future reliance on these technologies. 
Our work contextualizes the state of IP sharing, providing a uniquely broad perspective globally.

\end{abstract}

%% file: introduction.tex
\section{Introduction}

IP addresses are commonly shared across multiple devices and users, frequently due to 
 IPv4 address exhaustion. 
 In ISPs, this has included 
Network Address Translation (NAT) and Carrier-Grade NAT (CGN). 
However, IP sharing extends to other scenarios, including VPNs, proxies, enterprise gateways, and other browse-out systems, whether for anonymity, security, or facilitating access to geo-restricted content. 
While such IP sharing supports a variety of functional and operational purposes,
it also introduces problems, such as reducing the effectiveness of
IP-based attribution, blocklisting, rate limiting, and geolocation, and broadly
increases the complexity of network topologies~\cite{richter_multi_perspective_cgn, 10.1145/3419394.3423657}.
Furthermore, widespread CGN use can reinforce IPv4 dependence, slow IPv6 adoption, and add operational complexity, with uneven effects on regions with limited IPv4 resources. 

While many IPs are shared amongst a small number of users or devices (\eg a NAT
within a residential network or small business), some are shared at large scales
(\eg hundreds, thousands, or more users). These cases of large-scale sharing are
particularly important to identify and handle distinctly, especially with IP-based
mechanisms and policies, as they can exhibit unique behavior and the consequences of IP-based
actions can be substantially different. As a simple example, IP rate limiting
must account for the large number of users behind such addresses when
thresholding traffic rates, and mishandled decisions can cause widespread
collateral damage, interfering with many legitimate users and organizations.
Identifying these IPs based on traffic volume alone is unreliable, as other
highly active hosts, e.g., bots and web crawlers, also generate high traffic
volumes, as can malicious hosts. Using transport- or application-layer attributes
(e.g., port number distributions, User-Agent strings, or device
identifiers~\cite{cf_cgn_detection,richter_beyond_counting,casado_ip_based_client_ids_nsdi}) is limited to only certain vantage
points and offers mixed reliability, such as with minimal attribute diversity
exhibited by proxies that only relay specific applications or protocols~\cite{apple_private_relay, opera_browser_vpn, cloudflare_warp_split_tunnels}.
Overall, the manifestation of address sharing can differ substantially depending
on the underlying technology or deployment mechanism.

Although prior work has studied the extent of general IP address sharing on the Internet~\cite{richter_multi_perspective_cgn, livadariu_inferring_cgn, mandalari_natwatcher, mandalari2017trackingbignateurope}, no work has characterized the phenomenon of these particularly highly shared IP addresses across IPv4 and IPv6 networks at a global scale. 
Existing detection methods~\cite{richter_multi_perspective_cgn,
livadariu_inferring_cgn, mandalari_natwatcher,
mandalari2017trackingbignateurope} rely on client-based measurements or
properties of Peer-to-Peer traffic, yielding valuable insights but lacking the
global coverage needed to characterize address sharing at scale. 
 
In this work, we present a lightweight and more widely applicable approach to detect IPs that are shared at
large scales across many users, which we call \textbf{\massivelyshared IPs}. 
Our approach builds on an intuitive observation that the temporal properties of
network traffic volumes from such addresses become increasingly predictable once they reflect the aggregated activity of a sufficiently large population of end users and devices. 
In particular, traffic from an address shared among many subscribers smooths out
individual usage patterns, producing organic diurnal patterns. 
Building on this intuition, we present a method to formally detect these characteristics using Fourier analysis and curve-fitting evaluation. 
Since our method only requires traffic volumes per IP over time, it is lightweight and does not rely on any additional client-side information such as accessed content, packet-level specifics, or device identifiers that are otherwise used to identify sharing~\cite{cf_cgn_detection}. 
Thus, it can be used by a wide set of stakeholders in network operations, such
as transit Internet providers, those with encrypted traffic or flow data, and samples from high-volume links. 

We deploy our method across access logs of a major global Content Delivery
Network (CDN), providing an unprecedented characterization of \massivelyshared IP addressing across IPv4 and IPv6, spanning geographic regions, access networks, and cloud environments. 

Our key contributions include:

\begin{enumerate}[leftmargin=0.2in,topsep=0.01in]
	\item We develop a method to identify \massivelyshared IP addresses based on diurnal traffic and smoothness, informed by Fourier analysis of empirical patterns, and evaluate it on both synthesized and real-world shared traffic. We evaluate and characterize our method, including heuristics to estimate how many subscribers are behind \massivelyshared IPs, proper parameters, and suitable timeframes.

	\item We present a large-scale characterization of client IPv4
\massivesharing on the Internet, quantifying it globally and
regionally. We find that while only 1-3\% of active IPs are
\massivelyshared, they originate between 25-54\% of all IPv4 traffic, depending 
on the country. This demonstrates a stark concentration of traffic
from \massivelyshared IPv4 addresses.

	\item Broadly, we find 82\% of all \massivelyshared traffic originates from end-user ISPs. 
	When we observe \massivesharing in IPv4, we ask whether the networks also have IPv6 connectivity, finding that only 55\% of top providers with \massivelyshared IPv4 deploy IPv6. 
	We find that IPv6 sharing in end-user ISPs is rare, but sometimes deployed in cellular setups. 
	
	\item Notably, we find that 14\% of \massivelyshared traffic comes from cloud providers. We find that \massivelyshared cloud traffic is largely driven by proxy services~\cite{apple_private_relay,cloudflare_warp}.

	\item Finally, we evaluate \massivelyshared traffic over time, finding that the fraction of traffic from \massivelyshared IPv4 IPs steadily increases.

\end{enumerate}

%% file: related_work.tex
\section{On Shared Address Detection}
\label{sec:background}

\PP{Related Work}
Although no work has considered the traffic quantity of shared IPs, 
prior work has leveraged traffic features to identify and quantify users behind home NAT deployments, including using IP TTLs,  HTTP User-Agent strings~\cite{maier_nat_usage}, and BitTorrent client identifiers~\cite{livadariu_inferring_cgn}. 
Others have used IP leakage from BitTorrent~\cite{richter_multi_perspective_cgn} and on-client measurements to identify and characterize NAT deployments~\cite{richter_multi_perspective_cgn, mandalari_natwatcher, mandalari2017trackingbignateurope}. 
Others have used diversity in User-Agent strings~\cite{casado_ip_based_client_ids_nsdi} and source ports~\cite{cf_multi_user_ip_addr} to identify shared IP addresses. 
A CDN has extended this to identify CGN using ML classifiers over a variety of unspecified signals, including client diversity and traffic volume, then validated using PTR and WHOIS records that contain hints about address deployment~\cite{cf_cgn_detection}. 
Distinctly, we identify \massivelyshared IPs, focusing on high-traffic IPs that emit the majority of traffic. 

\PP{Challenges in Sharing Detection} 
Detection of massive IP sharing remains challenging because the necessary information about address sharing is rarely exposed, especially at scale. Approaches that rely on transport- or application-layer attributes to identify sharing require vantage points with such data available. 
While this data may be available to individual web platforms, transit providers cannot observe these attributes if they are encrypted. 
However, smaller web platforms cannot observe broad attributes such as port distributions, whereas an ISP or transit provider could. 
Moreover, different types of IP sharing setups may or may not result in detectable evidence. For example, proxy services such as Apple Private Relay only work for select applications~\cite{apple_private_relay, opera_browser_vpn, cloudflare_warp_split_tunnels}, so it may not result in a large number of distinct User-Agent strings~\cite{cf_multi_user_ip_addr}. Carrier-Grade NAT setups can use a plethora of configurations~\cite{richter_multi_perspective_cgn}, making detection based on transport-layer attributes, such as source ports, only partially effective. 
Thus, methods relying on these attributes can only observe a subset of IP sharing depending on the technology behind it. Finally, as we focus on \massivelyshared IPs, one may assume that particularly high volumes of traffic (e.g., client requests) may be sufficient to detect them. However, high volume alone cannot discriminate high numbers of end users from automated network traffic, such as crawling or infrastructure, as we will study in Section~\ref{sec:cloud}. 
Although an RFC published in May 2026 proposes a way for operators to report shared prefixes~\cite{rfc9977}, it is not clear if and how widely it will be adopted.

\PP{Overcoming Challenges Towards Global Characterization}
We overcome these challenges by detecting the traffic shape produced by \textit{many} users sharing an IP address: smooth and diurnal request patterns. 
Our method only requires traffic volume over time and is agnostic towards the actual implementations or technologies used to share IP addresses across users.
We elaborate on the specificities of our method in Section~\ref{sec:method}. 
Indeed, while our method is limited by user behavior and measurement vantage, other methods are limited by the technology responsible for IP sharing. 
As we will show, the CDN access logs provide us with a global view, allowing us to characterize \massivelyshared IPs, agnostic of the technology behind them.

%% file: data.tex
\section{Data Sources and Processing Steps}
\label{sec:data}
In this section, we introduce our vantage point and the corresponding datasets we use.

\subsection{Primary CDN Log Data}

\PP{Hourly Request Counts} 
The primary data source for our method and analysis is server logs of one of the largest global CDNs. In 2026, this CDN serves end-users globally by operating over 350,000 servers in over 1,600 networks in over 130 countries. Each time a client fetches a Web object from a CDN edge server, a log entry is created, which is then aggregated through a distributed data collection framework. We have access to the exact number of requests issued per hour from individual IPv4 addresses and for IPv6 /64s. 
We note that in IPv6, an entire /64 is typically assigned to an end-site~\cite{rfc6177}. 
We observe hourly request counts, or ``hits'', for IPv4 addresses and IPv6 /64s. 
We refer to an IP as ``active'' if the IPv4 address or IPv6 /64 has at least one successful request during the analysis period. 
For the bulk of our analysis, we focus on the week of July 13th, 2025. 
Later, we show long-term trends from January 2024 to October 2025.

\subsection{Contextual Datasets}

\PP{Identifying Cellular Clients}
We contextualize our dataset using the CDN's Real User Monitoring (RUM) system. 
The RUM system leverages the Network Information API~\cite{network_info_api} implemented in Android's browser, among others, and reports
whether the client is using a cellular or WiFi connection.
We follow prior work~\cite{rula_cell_spotting} to measure and classify prefixes (/24s in IPv4, /48s in IPv6) as cellular or non-cellular, including their treatment of mixed-access prefixes and the use of a 0.5 cellular-to-WiFi measurement ratio threshold for cellular classification. This threshold was shown to provide a conservative estimate of cellular networks.
Furthermore, to ensure representative data per network, we only use networks where we have at least 10 measurements. 
In total, we use 112M measurements to label some 2.5M IPv4 /24s and 1.4M IPv6 /48s address blocks. 

\PP{Measuring IPv6 Connectivity for IPv4 Networks}
Similar to prior work~\cite{padmanabhan_dynamips}, for dual-stack clients, 
we record their prefixes in each IP version (/24s in IPv4 and /48s in IPv6). 
When a client makes a request to a dual-stack domain hosted by the CDN, the server records the current client prefix (IPv6 /48 or IPv4 /24, but typically IPv6 because of happy eyeballs~\cite{sattler_lazy_eye_inspection}), then launches an additional Web request towards a domain that only supports IPv4 and records the /24. 
We use this to tag some 1.7M IPv4 /24s as also having IPv6 connectivity to contextualize deployments.

\PP{User-Agent data} 
We additionally log the User-Agent (UA) string present in a random sample of 0.1\% of client requests. We use this data to contextualize client traffic patterns in Section~\ref{sec:cloud}. 

\PP{Geolocating Data}
We geolocate all client IP addresses using the proprietary CDN geolocation database to country-level granularity. We point out that the Service Level Agreement (SLA) of the geolocation database covers 99\% country-level accuracy and point out that 
country-level geolocation for clients has been found to be consistent across different methods and datasets~\cite{2011-huffaker-gt, poese2011ip}.

\PP{Labeling Network Types}
Throughout our analysis, we label ASN types primarily using data from IPInfo~\cite{ipinfo} to classify ASNs into 5 categories: ISP, hosting, education, business, and government. 
To verify our labels, we manually review the labels of the top 10 providers in each of the top 10 countries per continent. 
If necessary, we change the labels by consulting with other data sources~\cite{peeringDB, bgp_tools} and looking at the websites of the company owning the ASN. 
Furthermore, we add labels for unlabeled ASNs until we cover over 99\% of traffic per continent. 

%% file: method.tex
\section{Method}
\label{sec:method}

In this section, we present  our method to detect \textit{diurnal} and \textit{smooth} patterns in client requests. 
We leverage the Fast Fourier Transform (FFT) to decompose a time series of requests per hour into frequency-domain components. We first show a motivating example, review related work on detecting diurnal network characteristics, and then introduce our method and parameter choices. 
We include a statement on ethics in Appendix~\ref{app:ethics}.

\begin{figure}
	\begin{subfigure}[]{.49\linewidth}
		\includegraphics[width=\linewidth]{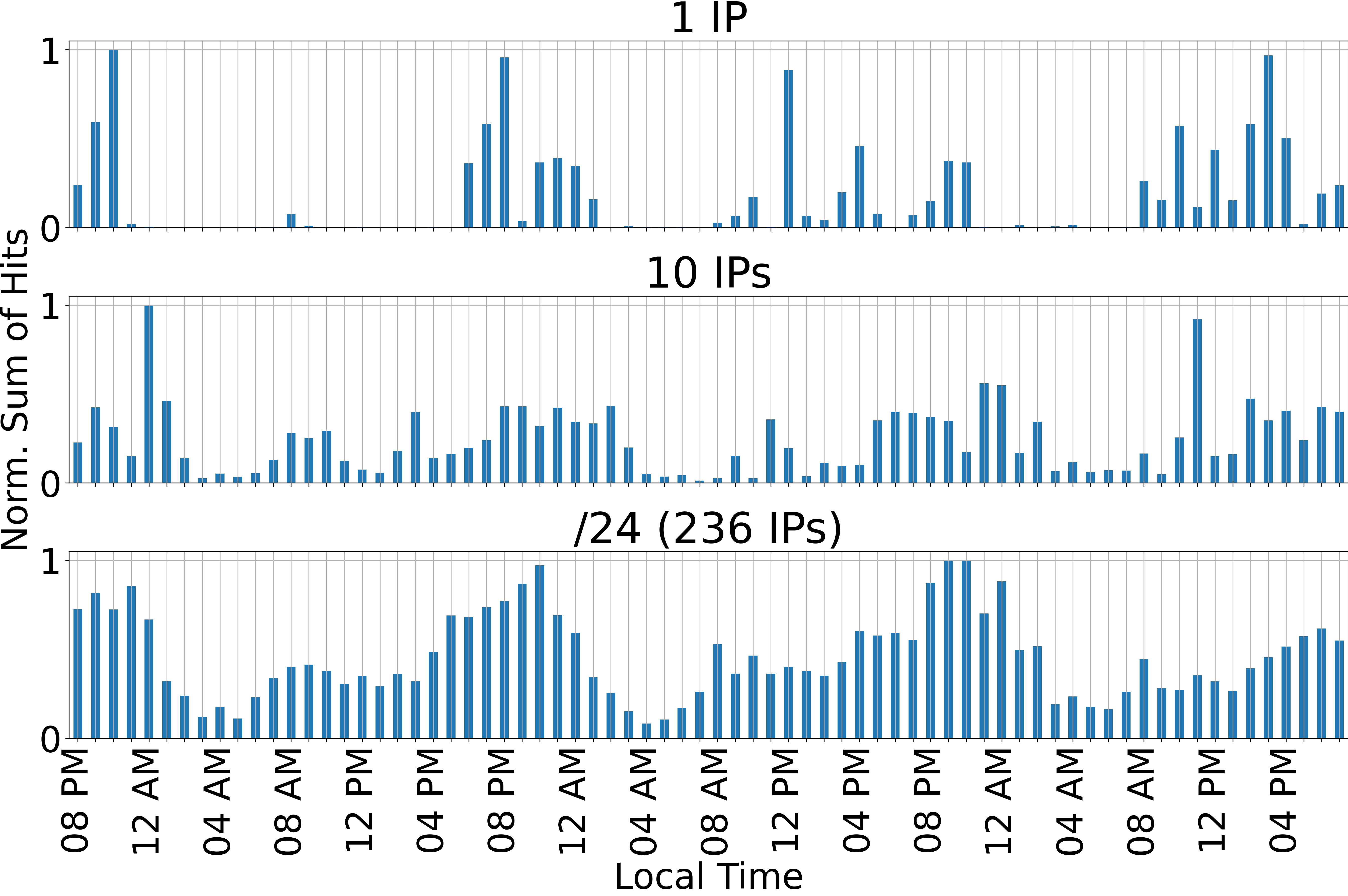}
		\caption{US broadband ISP A.}
		\label{fig:ip_agg_examples}
	\end{subfigure}
	\begin{subfigure}[]{.49\linewidth}
		\includegraphics[width=\linewidth]{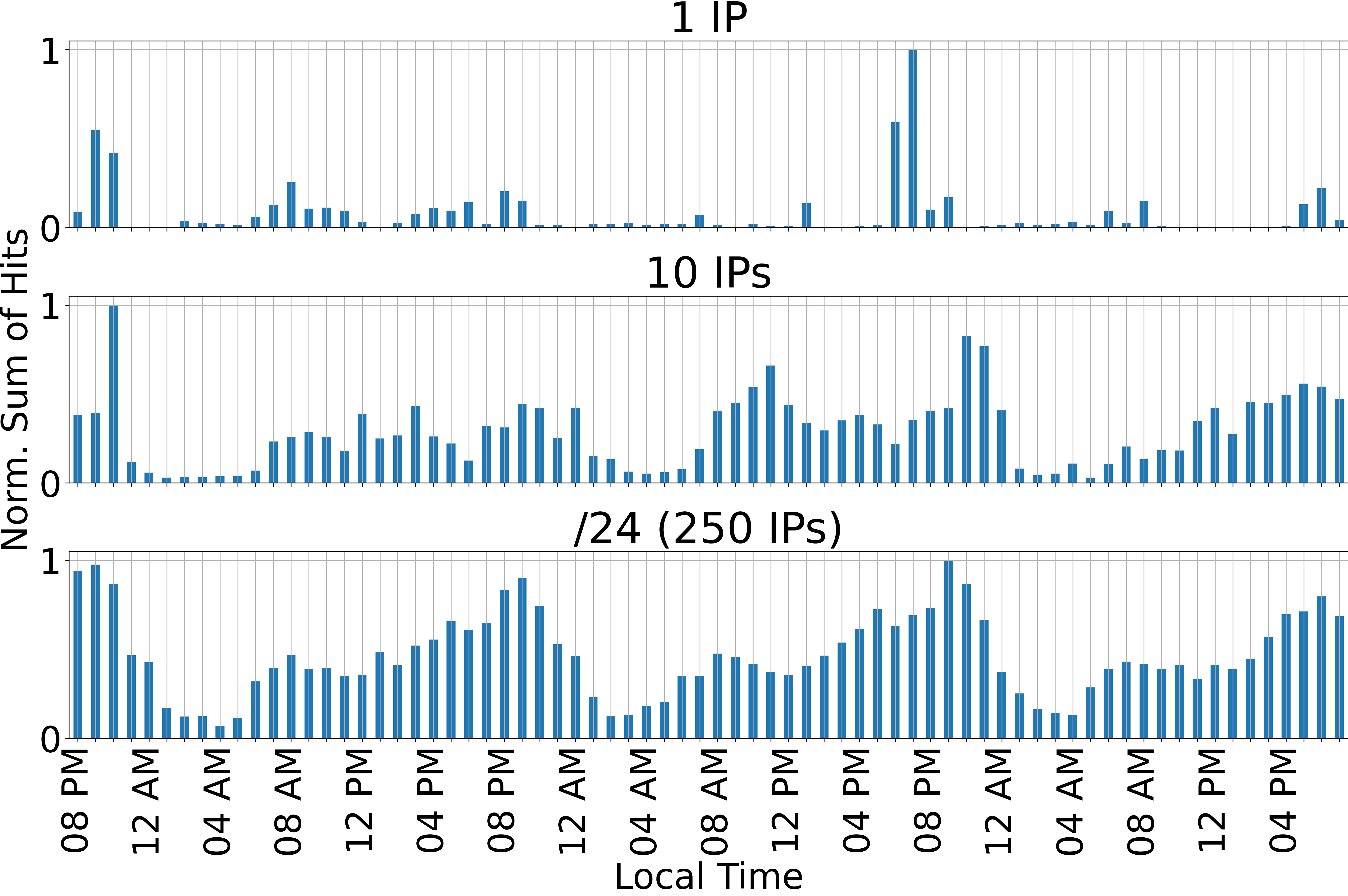}
		\caption{US broadband ISP B.}
		\label{fig:isp_b_ip_agg_examples}
	\end{subfigure}
	\caption{Traffic over 3 weekdays (72 hours) for different IP aggregate sizes within US ISPs A and B known to deploy single-subscriber IPv4. In each plot, we sum and normalize the number of hits from all IPs.} 
	\label{fig:us_isp_ip_agg_examples}
\end{figure}

\subsection{Motivating Example: Synthesizing Massively Shared Traffic} 
\label{sec:motivating_ex}
We first provide a motivating example to highlight the desired attributes our method detects.  
We exemplify what shared traffic looks like by \textit{synthesizing} what traffic from many subscribers looks like. 
To do this, we inspect traffic patterns of address blocks that connect end-users to the Internet and are known to assign a unique IPv4 IP per subscriber~\cite{att_deployment,comcast_deployment}. 
That is, these IPs may be shared locally, \ie within a household via NAT; however, they are not \textit{massively} shared, \eg as opposed to a CGN IP. 
We then show how the shape of requests changes when aggregating the traffic from individual IP addresses together. 
To illustrate why this is a strong indication of sharedness, we present motivating examples from two large US broadband providers. 
We collect these address blocks by examining probe tags on RIPE Atlas~\cite{ripe_atlas_platform} and by manual collection from colleagues who use these providers. 
Figure~\ref{fig:us_isp_ip_agg_examples} shows examples of traffic patterns from IPs in two IPv4 /24 networks.
In traffic patterns of individual IPv4 addresses, while some diurnal patterns emerge (\ie no traffic late at night), we see high variability. Aggregating traffic from 10 IPs within the same network starts to provide more diurnal visual clues, with peaks and dips in traffic at logical times. 
Finally, we see that predictable patterns emerge when we look at traffic from the entire /24.

Our method is centered around systematically detecting these resultant \textit{diurnal} and \textit{smooth} traffic patterns across traffic from individual IP addresses to detect \massivesharing. 
That is, on top of strong daily patterns, we seek traffic that is also predictable (\ie less variable). 
In the next sections, we will formalize our method, which allows us to detect these patterns at scale.

\subsection{Related Work: Detecting Diurnality}
\label{sec:related_work_diurnal}

Prior work has used FFT analysis to determine diurnality, specifically by analyzing whether 1/24 is the strongest frequency and its energy~\cite{quan_internet_sleeps,baltra_ebb_flow}. 
However, this definition was used in an analysis of the number of online IPs in a network in a given time period, aligning with common patterns in IP churn (\ie 24 hours). 
Because our data is based on traffic from IPs and therefore subject to different patterns and trends, 
we build upon this definition in Section~\ref{sec:method}. 
Other work has detected diurnal patterns in IPv6 traffic over entire ISPs~\cite{strowes_diurnal_ipv6_traffic},
 or identified diurnal IPv6 patterns over all traffic generated by 
individual households~\cite{valapu_non_binary_ipv6_adoption}. 
However, we highlight that the authors measured 5 residences and noted high variation in traffic patterns. 
Others have used diurnal patterns in traffic as an indication that traffic is not spoofed or a part of an attack~\cite{lichtblau_detection_spoofed_source_ips}, or more broadly to forecast backbone traffic~\cite{papagiannaki_forcasting_backbone_traffic} and anomaly detection~\cite{barford_signal_analysis_traffic_anomalies}.

\subsection{Detection Method and Criteria}
\label{sec:detection_method}

\PP{Fourier Analysis to Identify Diurnal Traffic} 
Following our observation on diurnal traffic shapes in Section~\ref{sec:motivating_ex} and prior work~\cite{baltra_ebb_flow,quan_internet_sleeps}, we use Fourier Analysis to quantify diurnality in traffic patterns. 
We perform a Fast Fourier Transform (FFT) on a time series of the number of "hits" per hour to define and reconstruct the periodic signal. 
Similarly to prior work, we consider traffic diurnal  
if 1/24 (a 24-hour period) is the frequency with the most energy. 

\begin{figure}[t]
    \centering

    \begin{minipage}{0.48\linewidth}
        \centering
        \includegraphics[width=\linewidth]{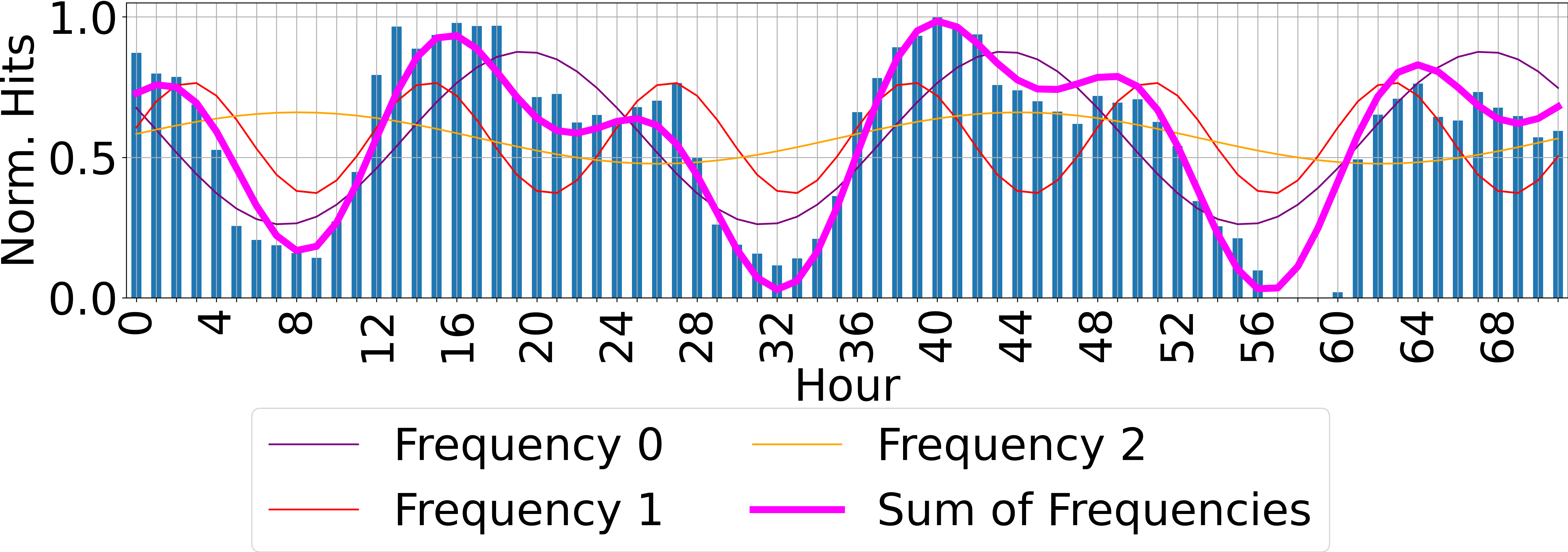}
		\caption{To illustrate how we use the FFT output to reconstruct the traffic pattern, we plot the top 3 strongest frequencies (centered at the mean of the hit volume to visually highlight how the curve fits the data) and their sum for the traffic from an IP in a proxy service. 
		Each sinusoidal function captures a periodic behavior in the traffic, and when added together, they fit the curve.}
		\label{fig:proxy_signal_reconstruction_freqs}
    \end{minipage}
    \hfill
    \begin{minipage}{0.48\linewidth}
        \centering
        \includegraphics[width=\linewidth]{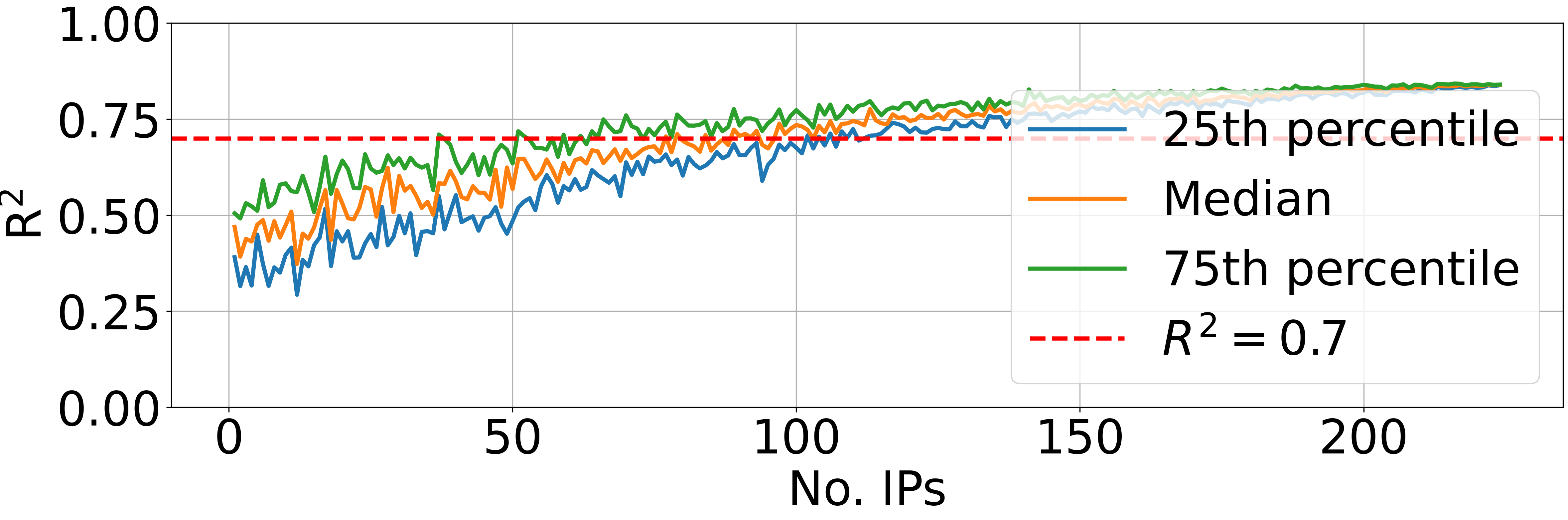}
	\caption{We experiment with traffic aggregates in a /24 from US ISP A. 
		Starting with 1, we randomly choose $N$ IPs 25 times, reconstruct the traffic with the top 3 frequencies from the FFT, 
		and compute the $R^{2}$ value between the reconstruction and the data. 
		We plot the $R^{2}$ values of each quartile for each experiment with $N$ IPs. 
		Broadly, the $R^{2}$ value (smoothness) increases 
		as we add more IPs (aggregate more subscribers).}
	\label{fig:r_squ_increase_example}
    \end{minipage}

\end{figure}

\PP{Modeling Smoothness} 
Based on our observations, we introduce one more criterion: smoothness. 
 We want to identify traffic that follows a relatively smooth sinusoidal signal without much randomness. 
An FFT decomposes a time series into sinusoidal curves with different strengths and frequencies. 
When all these are added together, the signal is reconstructed, and, for example, would match the traffic pattern of the single IP in Figure~\ref{fig:us_isp_ip_agg_examples}.
In general, when fewer frequencies are used, the match to the original signal degrades.
Our idea is that the notion of \textit{smoothness}
is when the match to the original signal remains good even when just a few of the strongest frequencies are used. 
That is, if only a few sinusoidal signals can reconstruct the shape of the traffic, it must be smooth, as it is not possible to reconstruct a chaotic shape with so few frequencies. 

After experimentation, we choose to use just the 3 strongest frequencies (plus the DC offset, the $y$-axis offset added to the reconstruction) to reconstruct the signal. We show the three top frequencies and their addition to reconstruct traffic from a shared IP using a proxy service as an example in Figure~\ref{fig:proxy_signal_reconstruction_freqs} as a visual example of how the FFT deconstructs the shape of the curve. 
Choosing this parameter is a trade-off, although we potentially have more false negatives but we have higher confidence that the IPs we do identify are \massivelyshared. 

\PP{Evaluating Curve Fit}
Then, we measure how well the reconstructed signal fits the time series by calculating the $R^{2}$ (also known as the coefficient of determination) as a measure of the variance captured and goodness of fit for our curve~\cite{Weisberg2014,ChatterjeeHadi2012}. 
The $R^{2}$ value is between 0 and 1, with higher values indicating a better fit, \ie better \textit{smoothness}.

To evaluate how the curve fits vary, we repeatedly aggregate a random set of $N$ IPs from a /24 in US ISP A and calculate the $R^{2}$ value from the reconstruction. 
We perform this from $N$=1 to the number of active IPs in the /24 (235 total) and present our findings in Figure~\ref{fig:r_squ_increase_example}. As we aggregate traffic from more IPs, $R^{2}$ increases and shows less variability, indicating a smoother fit and more uniform behavior for larger aggregates of subscribers. In this work, we choose our threshold for $R^{2}$ at $0.7$, which we find to be consistently exceeded once 75 subscribers/IPs are aggregated. 
We also note that $0.7$ is a commonly accepted value for a good curve fit~\cite{meainingful_r_sq}. We study the sensitivity of our choice in Appendix~\ref{sec:sensitivity}. 

\PP{Criteria for \MassivelyShared IP addresses} 
In this work, we label an IPv4 address, or an IPv6 /64,  as \massivelyshared if its respective request pattern meets two criteria: \textit{(i)} the strongest frequency is 1/24, and \textit{(ii)} when using the 3 strongest frequencies to reconstruct the signal using an FFT, the $R^{2}$ value between the data and the reconstruction is greater than $0.7$. This definition results in a binary classification of IPs as either \massivelyshared or \nonmassivelyshared. 
Practically speaking, we label an IP address as \massivelyshared if the hourly request pattern reaches a strongly diurnal pattern with a  degree of smoothness that matches our synthesized traces.

We evaluate how our findings change alongside different $R^{2}$ parameter values longitudinally, finding that results do change slightly with different values, but traffic quantities are consistent over time. 
We discuss the $R^{2}$ parameter sensitivity in more detail in Appendix~\ref{app:method_details}.

\PP{Implementation: Time Period}
We apply our method to 3 days of data per IP on Tuesday, Wednesday, and Thursday to avoid differences in diurnal traffic patterns due to weekend behavior (\eg for business networks) and differences in days of the week considered weekends globally. 
We acknowledge that identifying differences in traffic on weekends is interesting in itself, and discuss it along with other future directions in Section~\ref{sec:future_work}. 
Finally, we also acknowledge that IP churn in our analysis period could result in a poor fit.
We discuss this and other limitations in Section~\ref{sec:limitations}. 

 \begin{wrapfigure}{r}{0.5\textwidth}
	\includegraphics[width=\linewidth]{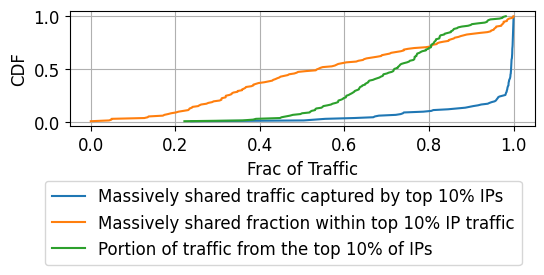}
	\caption{
			The traffic concentration per country and its intersection with the \massivelyshared IPs. 
		}
	\label{fig:frac_shared_top_10p_per_country}
 \end{wrapfigure}

\section{Method Evaluation}
In this section, we initially contextualize the IPs our method identifies. 
We estimate the number of subscribers behind the IPs, compare our identified \massivelyshared IPs with relative traffic volumes, and compare them to the IPs of several specific known cases. 
We clarify that we 
contextualize our broader characterization at a global scale in Sections~\ref{sec:sharing_overview}, \ref{sec:isps}, \ref{sec:cloud}, and  \ref{sec:sharing_over_time}.

\subsection{Estimating Usage Behind \MassivelyShared IPs} 
\label{sec:quantifying_usage} 
To estimate how many individual subscribers are behind a \massivelyshared IP, we aggregate traffic from individual IPs and analyze how many it takes to reach our \massivelyshared criteria, as presented in Figure~\ref{fig:r_squ_increase_example}. 
While we cannot claim to precisely quantify the number of users behind a \massivelyshared IP, 
 we use hits aggregated from individual subscribers 
 as a heuristic to estimate the user scale at which traffic begins to satisfy our \massivelyshared criteria. 
As shown in Figure~\ref{fig:r_squ_increase_example}, in a US ISP, we find that our criteria for \massivelyshared are consistently reached around 75 single-subscriber IPs. 
We repeat this experiment with other /24s in this US ISP and other networks with single subscriber IPv4 IPs in Germany, France, Japan, and Iran. 
We find that we reach the \massivelyshared criteria after aggregating traffic from 45-160 IPs. 
We discuss these analyses in more detail in Appendix~\ref{app:other_countries}. 
These values vary with CDN presence, browsing activity, and other network infrastructure, but indicate that the IPs we label are generally shared by \textit{at least} 50-100 subscribers. 
We emphasize that this number is a lower bound, as after this number is reached, synthesized traffic consistently meets our criteria for \massivelyshared. 
In Section~\ref{sec:characterizing_detected_ips}, we further characterize the IPs we are able to detect relative to the country's total traffic volume.

\subsection{\MassivelyShared Addresses and Traffic Volume}
\label{sec:characterizing_detected_ips}

We next compare \massivelyshared IPs with their traffic concentration per country. 
Specifically, in Figure~\ref{fig:frac_shared_top_10p_per_country}, we analyze the traffic from the top 10\% of IPs (ranked by volume of traffic sent). 
In 85\% of countries, we see that the top 10\% of IPs send over 90\% of all traffic from massively shared IPs per country. 
Thus, \massivelyshared IPs are typically among the highest-volume IPs per country. 
However, we emphasize that \textit{high volume alone is not enough} to detect \massivelyshared IPs. 
That is, in only 17\% of countries is \massivelyshared traffic more than 90\% of traffic from the top 10\% of IPs. 
Therefore, these high-traffic IPs are also comprised of infrastructure (\eg API calls), crawlers, or unusually high-usage individuals. 
We emphasize the relevance of these high-volume IPs, as the traffic from the top 10\% of IPs per country often accounts for the majority of overall traffic (over 60\% in 77\% of countries), making operator decisions around them particularly impactful. 

\subsection{Validating with Known Cases}
\label{sec:verification}
In this section, we showcase examples of different cases 
that we expect to be \massivelyshared or not \massivelyshared. 
While we do have expectations for these technologies broadly, we have no ground truth for \massivelyshared IPs, especially which addresses will be \massivelyshared at a given point in time. 
Thus, we present these examples to characterize their deployment and our method. 
We discuss additional manual validation of individual technologies in Appendix~\ref{app:small_validation}. 

\PP{Cellular Networks in Mobile-Reliant Countries}
We analyze three African countries where we observe high rates of mobile reliance and IPv4 scarcity, implying \massivesharing~\cite{ipv4_allocs_per_country, internetsociety_internet_africa_2015}: Mali, Algeria, and Ethiopia~\footnote{We infer that these countries are reliant on mobile connectivity due to reports of high cellular usage and, in some cases, low reported fixed broadband subscriptions~\cite{fixed_broadband_subscribers_per_country, digital_report_ethiopia, digital_report_algeria, digital_report_mali}.}. 
Both of these characteristics lead us to expect that cellular IPs would be massively shared due to high reliance and low numbers of IPs. 
Using the cellular identification methodology described in Section~\ref{sec:data}, we cross-check cellular IPs with \massivelyshared IPs. 
We find that in Mali, 100\% of cellular traffic is from \massivelyshared IPs. 
In Algeria and Ethiopia, deployment varies slightly; 87\% and 76\% of cellular traffic is from \massivelyshared IPs, respectively. 
Thus, our findings align with our expectations that high-traffic cellular IPs in these settings are shared by many users. 
We explore the role of cellular connectivity in more detail in Section~\ref{sec:isps:ipv4}. 

\PP{Anonymization Proxy Services} 
We use publicly available geofeeds to study the global networks of two proxy services: Proxy 1 and Proxy 2. 
Proxy 1 is a secure service for businesses. 
 Proxy 2 can be used on mobile devices or computers but only works on a few applications and requires a paid subscription. 
In each case, we analyze these deployments in countries where these proxies appear to have high usage, making it likely that they are \massivelyshared. 
The geofeeds for these proxy services contain thousands of prefixes in both IPv4 and IPv6. 
Thus, for this exercise, 
 we count the proxy as highly used in a country if over 100 proxy IPs are in the top 10\% of IPs for that country, indicating they are indeed heavily used relative to the country's overall traffic. 

For Proxy 1, we observe that traffic is highly concentrated in \massivelyshared IPs. 
In 79\% of countries with high usage, over 99\% of proxy traffic comes from \massivelyshared IPs, and in all countries with high usage, over 90\% of proxy traffic comes from \massivelyshared IPs. 
We hypothesize that most IPs from this service are \massivelyshared because entire business networks use this service for all or nearly all of their traffic, resulting in high amounts of users. 
Proxy 2 aligns with \massivelyshared IPs, although it varies more per country. 
In Proxy 2,  in 75\% of countries with high usage, over 70\% of proxy traffic comes from \massivelyshared IPs. 
Thus, although this service sends a high relative traffic volume per country, they are not necessarily always used by a high volume of users. 
Because this service requires a subscription and only works on certain applications, we hypothesize that actual usage and adoption vary. 
Thus, in both proxies, our expectation that their networks broadly encompass \massivelyshared IPs is confirmed.

\PP{IPv4 Deployments That Are Not \MassivelyShared}
To study false positives ``in the wild'', we evaluate against two /16s (with over 40,000 active IPs each) from single-subscriber IPv4 deployments in ISP A. 
Less than 1-2\% of IPs are labeled as \massivelyshared. 
We additionally sample from 100 RIPE Atlas~\cite{ripe_atlas_platform} probes across 30 countries tagged as native IPv4 and active. 
Our method does not detect any of these IPs as \massivelyshared. 
Expanding our analysis to each IP in the /24 of the probes (over 11,000 IPs total), 2\% are \massivelyshared. 
We emphasize that, although we do not have ground truth for these networks (some IPs may indeed have a multitude of users, \eg in a business network or public WiFi, or improperly labeled probes), this result lends itself to the conservative nature of our method. That is, we favor false negatives over false positives.

\subsection{Limitations}
\label{sec:limitations}
We are foremost limited to the CDN's vantage and usage regionally. While we tested our method empirically in multiple countries, we may underestimate \massivelyshared traffic in regions with relatively fewer customers or users, resulting in traffic shapes that do not aggregate smoothly. In such cases, higher degrees of address sharing may be required to meet our criteria.

Our method is conservative by design. 
As discussed in  Section~\ref{sec:detection_method}, 
our method has trade-offs, but we purposely favor false negatives over false positives in order to report lower bounds on the global use of \massivelyshared IPs and their traffic. 
In other scenarios, the parameters of our method may be tuned in the opposite direction as appropriate. 
However, in this work, we may not detect all \massivelyshared IPs; for example, in Figure~\ref{fig:r_squ_increase_example}, we show that there is a chance that we do not detect traffic shared across high numbers of individual subscriber IPs. 
Another case that may impede our detection is IP churn in the middle of our analysis period, as mentioned in Section~\ref{sec:detection_method}, which could lead to a poor fit and mislabeled IPs. 
However, this is dependent on specific network properties and may happen within \textit{any} time period we choose. 
Thus, we instead optimize our time period for broad factors, such as weekends and workdays. 
In the opposite direction, we still may mislabel non-\massivelyshared IPs if they exhibit the traffic characteristics we observe, but Section~\ref{sec:verification} supports that such cases are generally rare.

Finally, we cannot quantify users behind \massivelyshared IPs precisely; we can only estimate subscriber counts with our heuristics to synthesize traffic, as patterns from individual users vary  and are dependent on deployment characteristics. 
Thus, we have limited vantage as to what is \textit{actually} behind \massivelyshared IPs, but have strong reason to infer that this is many subscribers. 

%% file: results.tex
\section{A Global View of \MassivelyShared IPs}
\label{sec:sharing_overview}
In this section, we analyze the results of deploying our method over the entirety of the IPv4 and IPv6 traffic from the CDN's access logs.

\PP{Overall Prevalence}
Globally, we find that 41.1\% of IPv4 traffic (CDN client requests) comes from \massivelyshared IPs, meaning that \textit{close to half of today's IPv4 traffic served by a CDN flows via \massivelyshared address infrastructure}. In terms of IPv4 addresses, we find that only 1.6\% of active IPv4 addresses are \massivelyshared, highlighting a heavily skewed distribution of traffic over the IPv4 space. 
In IPv6, we find that some 5.6\% of IPv6 client traffic comes from \massivelyshared\textit{ }/64s, and that these account for 0.35\% of active /64s during our measurement window. We are surprised to find substantial (though much lower than in IPv4) address sharing in IPv6, since IPv6 \massivesharing cannot be driven by address scarcity.

\begin{figure}[t]
	\centering
	\begin{minipage}{0.48\linewidth}
		\centering
		\includegraphics[width=\linewidth]{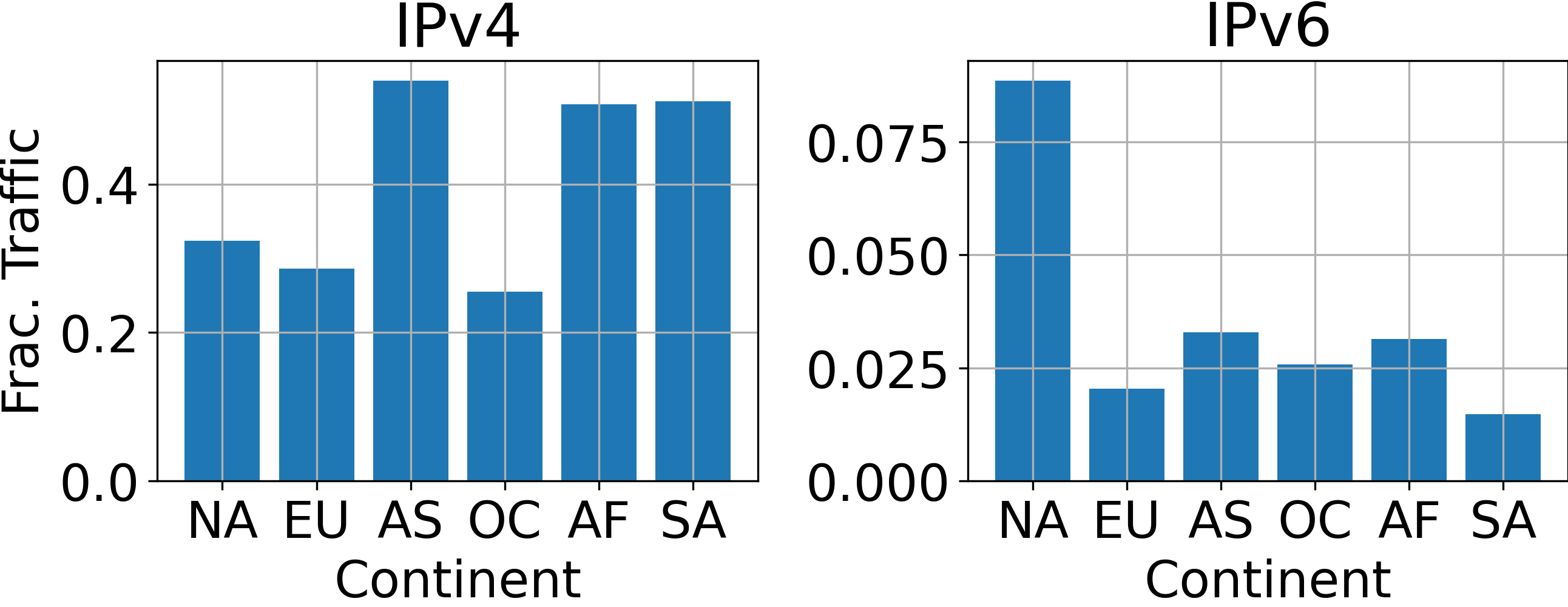}
		\caption{The fraction of traffic from \massivelyshared IPs per continent. Across all continents, IPv4 \massivesharing is pervasive, and particularly pronounced in Africa, Asia, and South America. IPv6 \massivesharing, while present, is less widely deployed.}
		\label{fig:all_asn_overview}
	\end{minipage}
	\hfill
	\begin{minipage}{0.48\linewidth}
		\centering
		\includegraphics[width=\linewidth]{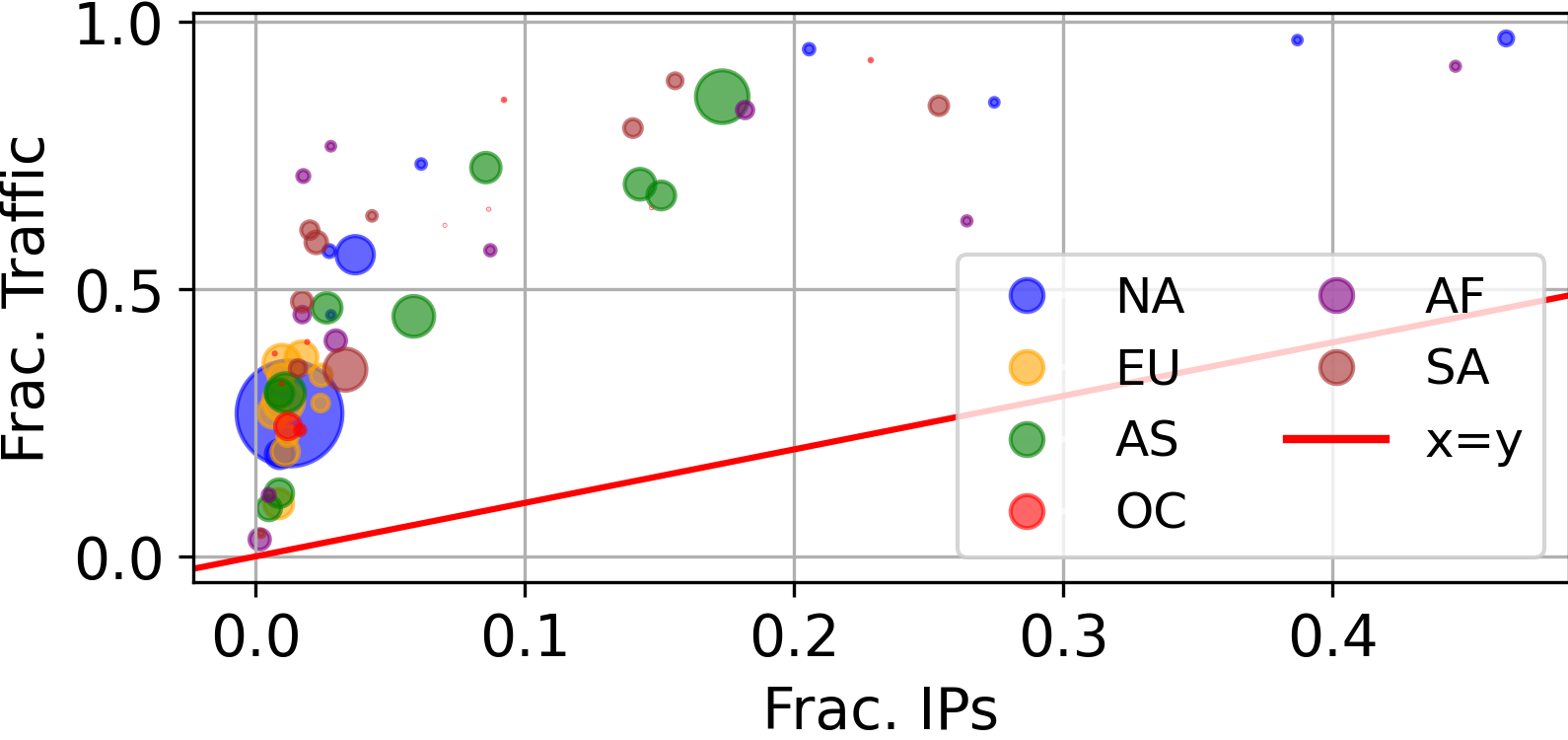}
		\caption{
			Traffic-to-IP fraction ratio for \massivelyshared IPs out of all active IPs and traffic per country. 
			We show the top 10 countries with the most traffic per continent. 
			The marker size is relative to the number of requests the CDN receives from the country. 
		}
		\label{fig:ipv4_shared_traffic_ratio}
	\end{minipage}
\end{figure}

\PP{Regional Prevalence}
Next, we present the fraction of \massivelyshared traffic per continent in Figure~\ref{fig:all_asn_overview}. 
In IPv4, we observe that more than half of client traffic seen in Asia, Africa, and Oceania comes from \massivelyshared IP addresses, with \massivesharing in Europe and North America slightly less prevalent at 29\% and 32\%, respectively. We point out that North America, especially the USA, has high volumes of IPv4 addresses, and Europe also has a large fraction of IPv4 addresses relative to population size~\cite{ip2location_ipv4_stats}, alleviating the need to share addresses widely. 
In IPv6, we observe that this is not the case, and IPv6 \massivesharing represents a much smaller fraction of traffic in each case. Notably, we find that North America dominates \massivelyshared IPv6 traffic (at 9\%).
Once again, we emphasize that while simply sharing IPv4 IPs is somewhat expected, our results highlight the scale at which a high number of users depend on \massivelyshared addressing.

\begin{wrapfigure}{r}{0.5\textwidth}
        \includegraphics[width=\linewidth]{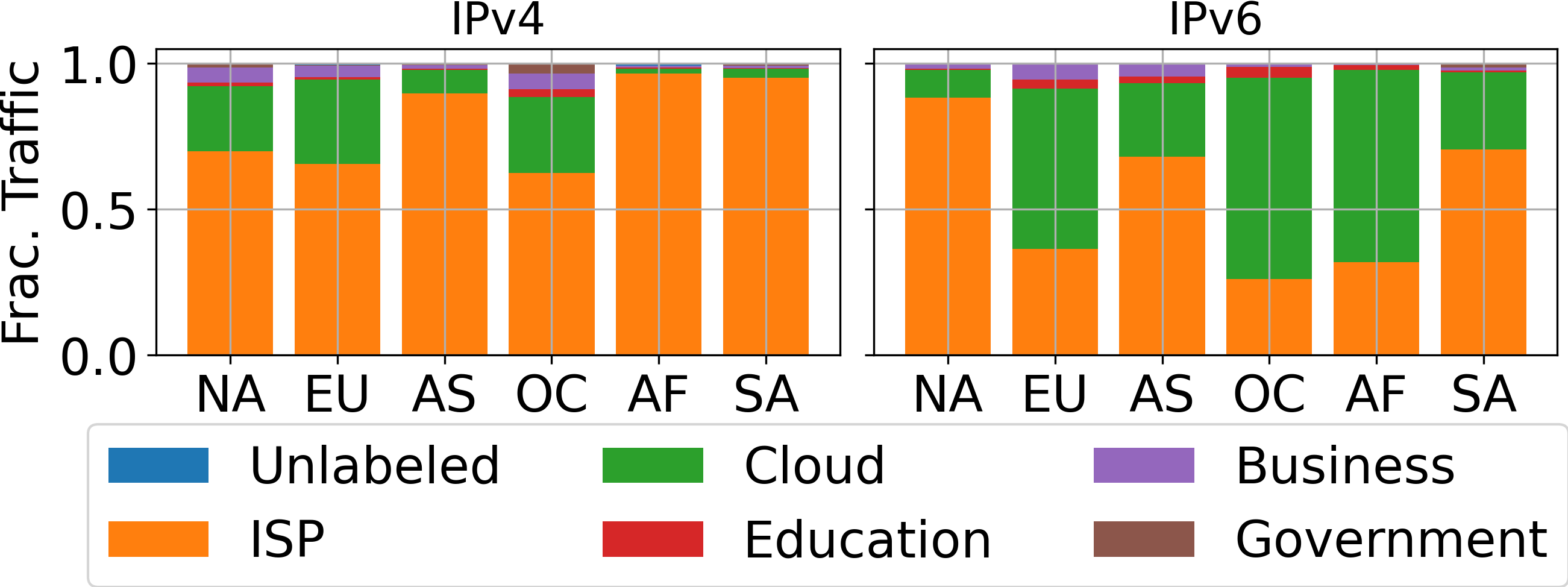}
		\caption{
		The distribution of \massivelyshared traffic among different AS types, per continent. 	
		\Massivelyshared traffic is dominated by ISPs and cloud providers across all continents. Recall that the \massivelyshared IPv6 traffic shown is a small fraction of total IPv6 traffic.}
		\label{fig:asn_di_traffic_categories}
\end{wrapfigure}

\PP{Country View of Traffic Concentration}
Next, we analyze \massivelyshared traffic at the country level. 
Although we already know from Section~\ref{sec:characterizing_detected_ips} that traffic per country is typically concentrated in a small fraction of active IPs, we now observe the magnitude of this concentration in \massivelyshared IPs. 
Given that address scarcity, technological deployment, and usage differ across regions, we complement our metric of \massivelyshared traffic fraction with the proportion of \massivelyshared IP addresses among all active IPs within each country. 
Figure~\ref{fig:ipv4_shared_traffic_ratio} shows this ratio for the top-10 traffic-ranked countries on each continent. 
In all cases, the fraction of \massivelyshared traffic is significantly higher than the fraction of IPs. While all selected European countries, as well as the USA, exhibit a small fraction of \massivelyshared IPs and shared traffic between 20\% and 40\%, many other nations show traffic ratios exceeding 50\%, and in some cases even 80\%. 
Specifically, in North America, traffic from Guatemala and Nicaragua is mostly from \massivelyshared IPs that encompass high fractions of all active IPs.
This ratio is 97\% of traffic from 44\% of IPs and 97\% of traffic from 39\% of IPs,  
 respectively. 
 Similarly, 
 in Africa, 97\% of traffic from Somalia originates from \massivelyshared IPs that encompass 62\% of all active IPs. 
 With high amounts of all active IPs being \massivelyshared and sending nearly all of the traffic for each country, 
 these findings align with prior observations that  limited Internet infrastructure and later access to IPv4 resources result in heavy reliance on CGN~\cite{richter_multi_perspective_cgn}.

In IPv6, \massivelyshared  /64s produce far less traffic. 
We find that 88\% of the countries we analyze have less than 3\% of their traffic from \massivelyshared IPv6 /64s. 
However, we highlight notable outliers. The United States has one major broadband and cellular provider that sends 47\% of its traffic from \massivelyshared IPs. 
Hong Kong has 43\% of its traffic from \massivelyshared IPs. Upon closer inspection, we observe that most of this traffic comes from cloud providers. 
We elaborate on this and other more granular observations on IPv6 sharing in Sections~\ref{sec:isps:ipv6} and~\ref{sec:cloud}.

\PP{Traffic By Network Type}
Next, we categorize ASNs by type and observe the amount of \massivelyshared traffic the CDN receives from each. In Figure~\ref{fig:asn_di_traffic_categories}, we find that the bulk of \massivelyshared traffic comes from End-User ISPs (82\% of all \massivelyshared traffic in IPv4, 80\% in IPv6) and cloud providers (12\% in IPv4, 16\% in IPv6). While this is intuitive for End-User ISPs, which must connect many users over a limited address space, we emphasize the scale that our findings of \massivelyshared IPs show. 
Moreover, we are surprised to observe a substantial portion of \massivelyshared traffic coming from cloud networks, a finding examined further in Section~\ref{sec:cloud}.

\begin{tcolorbox}[mytakeaway]
	\textbf{Takeaway:} 
	\Massivelyshared traffic compromises 41\% of the CDN's IPv4 client requests, but less than 2\% of active IPv4 addresses. Although present in all countries, \massivesharing is particularly pronounced in areas with later Internet deployment. On the other hand, we find just over 5\% of IPv6 traffic coming from \massivelyshared $ $ /64s. 
\end{tcolorbox}

\section{\MassivelyShared Addresses in End-User ISPs}
\label{sec:isps}

As discussed in Section~\ref{sec:sharing_overview}, most \massivelyshared traffic originates from End-User ISPs. Next, we focus on End-User networks, their use of massive sharing in IPv4, the presence of IPv6 capabilities, and unexpected cases of massive sharing in IPv6.

\subsection{\MassivelyShared IPv4 in ISPs}
\label{sec:isps:ipv4}

End-User ISPs present a natural deployment setting for \massivelyshared IPv4 addressing, given the need to connect large numbers of end users (``eyeballs''). 
As shown earlier in Figure~\ref{fig:all_asn_overview}, we find that Asia, Africa, and South America show particularly high levels of \massivelyshared addressing deployments in IPv4. 
This trend directly extends to ISPs, where 54\%, 51\%, and 50\% of all traffic originates from \massivelyshared IPv4 addresses in Asia, Africa, and South America, respectively, compared to 29\%, 23\%, and 20\% in North America, Europe, and Oceania.

\PP{Cellular Networks and \MassiveSharing}
Prior work has shown that IPv4 address sharing is ubiquitous in cellular networks~\cite{richter_multi_perspective_cgn}. 
Next, we ask what fraction of \massivelyshared IPv4 traffic in End-User ISPs is driven by cellular deployments, using context from the RUM measurements described in Section~\ref{sec:data} to identify /24s that are cellular. 
We note that this dataset contextualizes over 90\% of \massivelyshared IPv4 IPs as either WiFi or cellular, providing a wide coverage of cellular network labels.  
We find that the fraction of \massivelyshared ISP traffic from cellular connections varies by continent, with 26\% in North America, 52\% in Europe, 36\% in Asia, 10\% in Oceania, 66\% in Africa, and 20\% in South America. 
Thus, cellular plays a significant role in \massive IP sharing, but not all. 
Outside of Africa and Oceania, half or more of \massivelyshared IPv4 traffic from ISPs is from non-cellular deployments, likely with CGN. 
However, in Africa and Oceania, 7/10 and 6/10 of the countries, respectively, with the most \massivelyshared traffic have over 60\% of their \massivelyshared traffic from cellular networks. 
In the most extreme cases, Ethiopia's IPv4 traffic is 92\% \massivelyshared, and 70\% of that traffic is cellular; in Nigeria this is 83\% and 92\%, in Papua New Guinea this is 93\% and 91\%, and in Fiji this is 85\% and 84\%. 

\begin{tcolorbox}[mytakeaway]
\textbf{Takeaway:} \Massivelyshared IPv4 is deployed widely in broadband and cellular providers.  Although cellular is responsible for most \massivelyshared IPv4 traffic in Africa, implying a broader dependence on this type of connectivity, in other continents, the majority of \massivelyshared IPv4 traffic originates from non-cellular deployments.   
\end{tcolorbox}

\subsection{\MassivelyShared IPv4 and IPv6 Connectivity}
\label{sec:isps:ipv6_capabilities}

Next, we ask whether \massivelyshared IPv4 deployments also have IPv6 connectivity (\ie dual stack), indicating they have a path away from dependence on IPv4. 
Alternatively, standalone \massivelyshared IPv4 deployments may prolong IPv4 use, increasing reliance on exhausted address space while leaving networks misaligned with growing global IPv6 adoption~\cite{google_ipv6_stat}. 

\PP{Measuring IPv6 Connectivity}
For \massivelyshared IPv4 addresses, we are interested in whether the respective clients also have IPv6 connectivity. 
For this, we use the IPv4-IPv6 association data described in Section~\ref{sec:data}. 
We label a \massivelyshared IPv4 address as IPv6-capable if the respective /24 has an IPv6 association, labeling over 96\% of \massivelyshared IPv4 addresses. 

\begin{figure}
	\begin{subfigure}[]{.48\linewidth}
		\includegraphics[width=\linewidth]{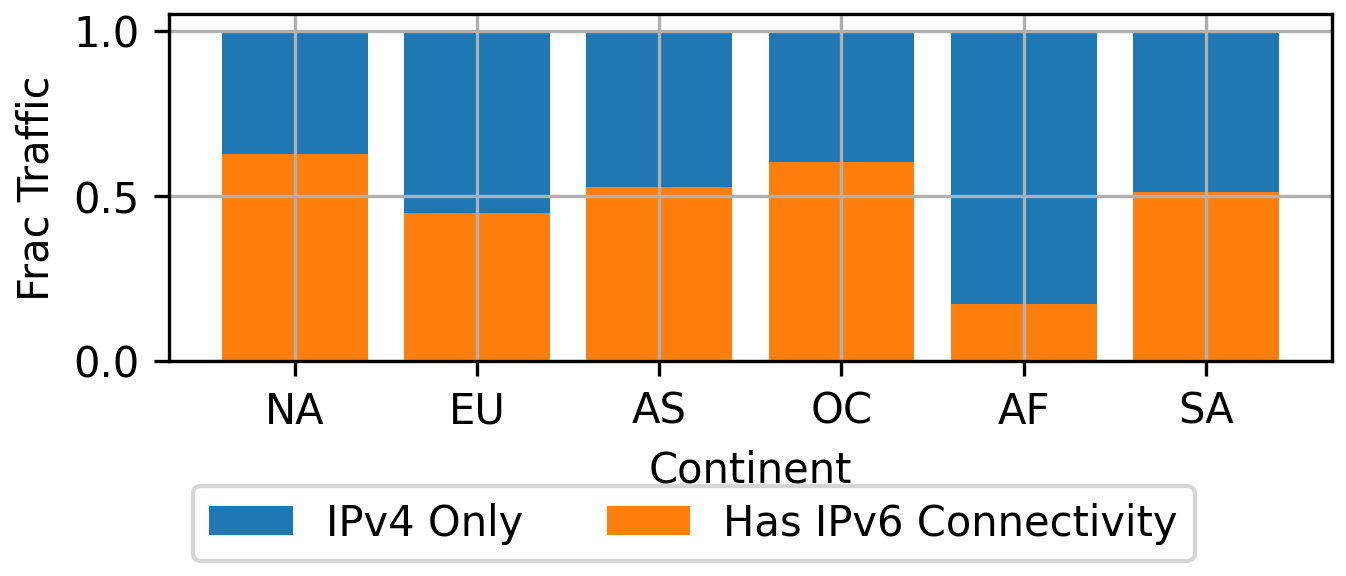}
		\caption{The fraction of \massivelyshared IPv4 ISP traffic from clients with IPv6 connectivity. 
		Across continents, many client networks lack IPv6 capabilities. 
		Africa has exceptionally low deployment, where many who share IPv4 have no IPv6 connectivity. 
		}
		\label{fig:ipv4_ipv6_assoc_shared_traffic_frac_isp}
	\end{subfigure}
	\begin{subfigure}[]{.48\linewidth}
		\includegraphics[width=\linewidth]{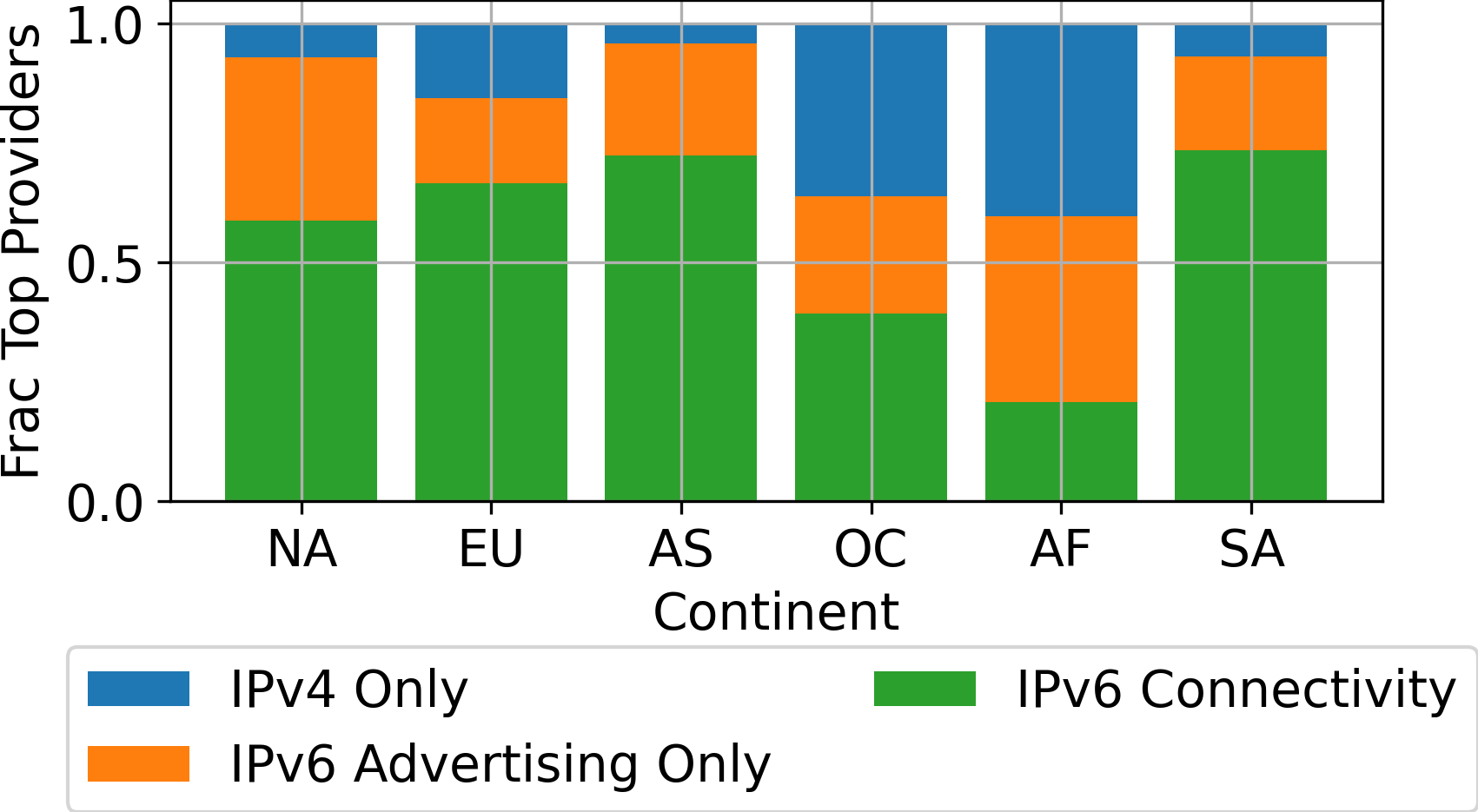}
		\caption{The IPv6 capabilities of the top-10 providers per country on each continent with the most \massivelyshared IPv4 traffic. 
		 }
		\label{fig:ipv4_ipv6_assoc_isp_asn_frac}
	\end{subfigure}
	\caption{An overview of IPv6 capabilities of the networks with associated \massivelyshared IPv4 traffic. We first provide an overview of IPv6 capabilities by amount of traffic, then by provider, including how many advertise any IPv6 BGP prefixes. We highlight that Africa has the highest fraction of IPv4-only providers, further characterizing this region as IPv4-dependent, as it also has the highest rate of \massivelyshared IPv4. }
	\label{fig:ipv4_ipv6_assoc_overview}
\end{figure}

Figure~\ref{fig:ipv4_ipv6_assoc_shared_traffic_frac_isp} shows the fraction of \massivelyshared IPv4 ISP traffic from IP addresses that have IPv6 connectivity. 
\footnote{Note that several factors can affect whether objects get fetched via IPv4 or IPv6 from the CDN, including CDN mapping considerations, customer and client-specific configurations, and network conditions~\cite{valapu_non_binary_ipv6_adoption, sattler_lazy_eye_inspection}.} At a high level, we find that about half of the \massivelyshared IPv4 traffic comes from IPs with IPv6 connectivity. Africa is the exception, with less than 20\% of the \massivelyshared traffic coming from dual-stack clients. Recall from Figure~\ref{fig:all_asn_overview} that Africa also has the highest fraction of \massivelyshared traffic overall. 
Our findings indicate that while \massivelyshared IPv4 networks are commonly dual-stack~\cite{ds_lite_rfc}, it is also often deployed as a standalone solution. 

\PP{A Multi-Dimensional View on IPv6 Deployment} 
Next, we analyze the degree of IPv6 deployment we observe in \massivelyshared IPv4 networks at the AS-level. 
In Figure~\ref{fig:ipv4_ipv6_assoc_isp_asn_frac}, we categorize the top 10 networks per country (in the top 10 countries per continent) in three categories: 
~\textit{(i)} ~IPv4 only, where we see \massivesharing in IPv4 and no indication of IPv6; 
\textit{(ii)} \textit{IPv6 Connectivity}: networks where we see \massivesharing in IPv4 and direct IPv6 associations; 
\textit{(iii)} \textit{IPv6 Advertising only}: networks where we do not measure
an IPv6 association, but we do
see that the networks announce \textit{any} IPv6 prefix in BGP. 
We use the CAIDA Pfx2AS dataset, built from Routeviews BGP data, to identify ASes that originate IPv6 BGP prefixes~\cite{routeviewsprefix2as}.

\PP{AS-Level Deployment Strategies}
As per Figure~\ref{fig:ipv4_ipv6_assoc_isp_asn_frac}, more than half of ISPs, in most regions of the world (Africa being the exception), do deploy \massivelyshared IPv4 and IPv6 that is not \massivelyshared. 
Specifically, 52\% of all providers we analyze use this deployment when classified by traffic. 
Although we do not explicitly detect single-subscriber or native IP deployments, we highlight that a common
 dual-stack deployment when IPvX to IPvY transition mechanisms are in place~\cite{nat64_rfc, 464XLAT_rfc, ds_lite_rfc} uses shared IPv4 and native IPv6. 
Our findings imply that these technologies \textit{may} be deployed widely across networks with \massivesharing.  
We elaborate on \massivelyshared IPv6 in Section~\ref{sec:isps:ipv6}.  

We highlight the importance of the differences between analyzing traffic and capabilities by provider. 
Although our results in Figure~\ref{fig:ipv4_ipv6_assoc_shared_traffic_frac_isp} largely align with other IPv6 deployment reports~\cite{apnic_ipv6_stat, google_ipv6_stat}, our per-provider results in Figure~\ref{fig:ipv4_ipv6_assoc_isp_asn_frac} paint a more complex picture. 
Looking at the top 10 providers per country for the top 10 countries per continent (100 total), 
IPv6 connectivity is commonly available at some point in the network, but we commonly see in Figure~\ref{fig:ipv4_ipv6_assoc_shared_traffic_frac_isp} that a large 
 quantity of \massivelyshared IPv4 traffic has no IPv6 associations. 
 This finding in large providers is impactful as they serve a significant part of the country's population, suggesting that IPv6 deployment metrics can overstate the extent to which end users behind \massivelyshared IPv4 infrastructure have IPv6 connectivity. 
Although we focus on large providers here, where we expect more IPv6 deployment, we see 79\% and 61\% of top ISPs in Africa and Oceania use IPv4 only (by connectivity). 
Importantly, we also find a sizable amount of ISPs globally that advertise at least one IPv6 BGP prefix, but no measurable IPv6 connectivity from \massivelyshared IPv4 clients. That is, a provider may deploy IPv6 unevenly within the network,  be making early attempts at configuring and setting up IPv6, advertise IPv6 as a part of an agreement with transit providers, or deployment may simply be a work in progress.

\begin{tcolorbox}[mytakeaway]
\textbf{Takeaway:} 	In 52\% of providers and in 50\% of traffic, dual-stack \massivelyshared IPv4 clients have \nonmassivelyshared IPv6.
We also find significant IPv4-only sharing deployments, particularly concentrated in Africa, where up to 79\% of top providers have no IPv6 clients.
\end{tcolorbox}

\subsection{\MassivelyShared IPv6 in ISPs}
\label{sec:isps:ipv6}

While IPv4 shows widespread \massivelyshared addressing, likely driven by scarcity, we unexpectedly observe that some ISPs also employ \massivelyshared address mechanisms in IPv6. Looking at \massivesharing in IPv6 ISPs, we find that North American (8\% total IPv6 traffic) and Asian ISPs (2\% total IPv6 traffic) are slightly higher than other continents.

\PP{\MassivelyShared IPv6 Cellular Traffic}
Toward understanding IPv6 \massivesharing, we analyze the fraction of cellular traffic among \massivelyshared IPv6 traffic. In Europe, Africa, and South America, cellular traffic represents less than 3\% of \massivelyshared IPv6 traffic, and in Oceania, it is just 10\%. However, in North America and Asia, cellular traffic represents the majority of \massivelyshared IPv6 traffic (92\% and 69\%, respectively). In North America, we note again that this phenomenon is due to one large provider (over two sibling ASes) that appears to share /64 prefixes across multiple subscribers. Within each AS, over 98\% of \massivelyshared traffic comes from cellular networks. 
We also highlight that this aligns with findings from prior work that analyzed end-user IPv6 addressing, which found similar behavior in the same ISP, finding \textit{over 10,000 users in a single /112}~\cite{li_end_user_ipv6}. 
In Asia, we also find that this is due to one large telecom provider in Indonesia, which is responsible for almost 70\% of all \massivelyshared IPv6 traffic in Asia. 

Although uncommon, IPv6 \massivesharing by large ISPs and cellular providers can generate substantial traffic. We hypothesize that providers may deploy it to maintain symmetry with IPv4~\cite{ipv6_nat_infobox}. If such practices become widespread, IPv6 may inherit shared IPv4's challenges.

\begin{tcolorbox}[mytakeaway]
\textbf{Takeaway:} While \massivelyshared IPv6 is not common in ISPs, we find surprising exceptions among large mobile providers who appear to share IPv6 /64s among multiple subscribers. If IPv6 sharing practices spread, known issues from the IPv4 sharing space may surface in IPv6 as well.
\end{tcolorbox}

%% file: cloud_results.tex
\section{\MassivelyShared Addresses in the Cloud}
\label{sec:cloud}

Although we find the majority of \massivelyshared traffic originating from End-User ISPs, 
we also find 14-16\% of \massivelyshared traffic (in IPv4 and IPv6, respectively) from cloud providers. We ask the question: What is this type of traffic, and are end users behind these IP addresses? 

\begin{figure}
	\begin{subfigure}[]{.49\linewidth}
		\includegraphics[width=\linewidth]{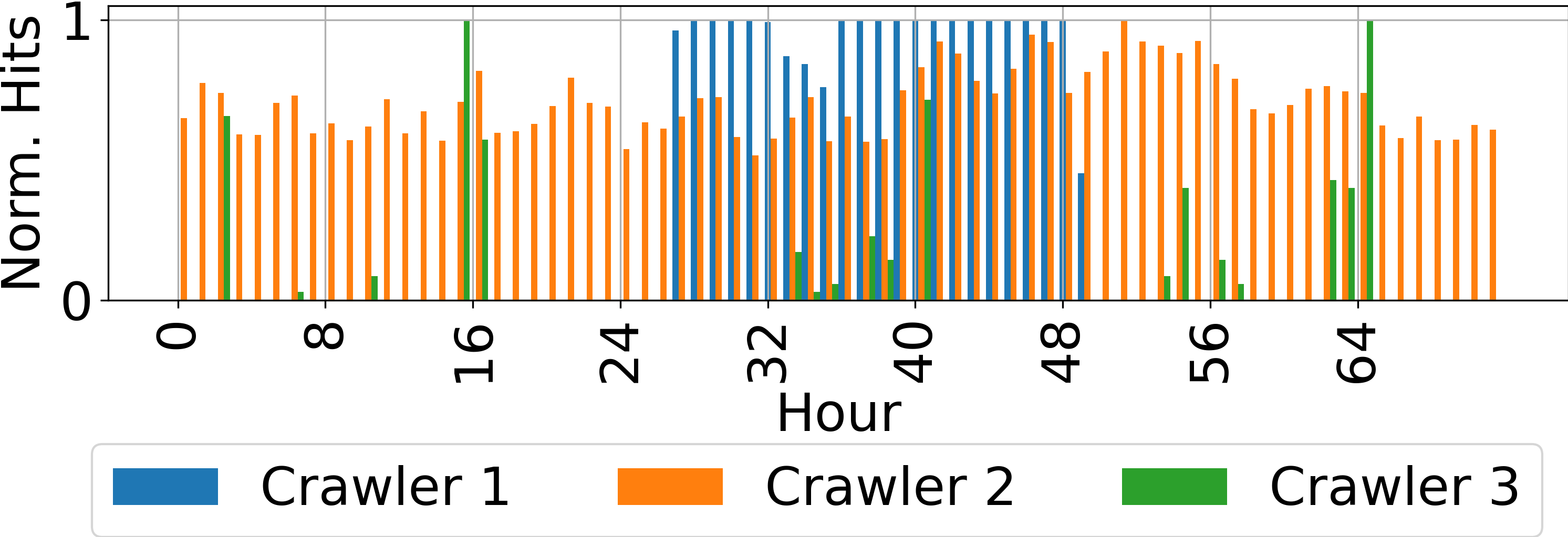}
		\caption{Traffic from bots on IPs that are not shared. 
		These bots self-identify as web crawlers for search engines. 
		We observe different traffic patterns, with some constantly crawling web services hosted by the CDN, and others only presenting sporadic traffic in bursts.}
		\label{fig:not_shared_bot_traffic}
	\end{subfigure}
	\begin{subfigure}[]{.49\linewidth}
		\includegraphics[width=\linewidth]{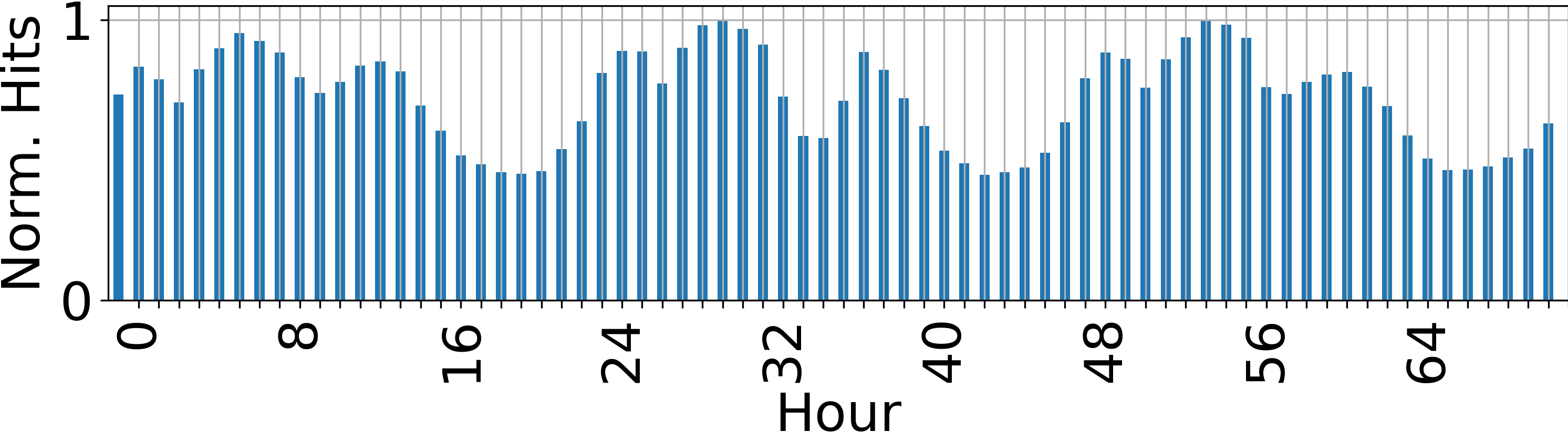}
		\caption{Traffic from a popular AI search engine. Searches are initiated \textit{on demand} when users query the AI. 
			This traffic meets our \massivelyshared criteria, 
			where although the end users are not (directly) behind the IP, the traffic is triggered by their actions. }
		\label{fig:diurnal_search_engine_bot_traffic}
	\end{subfigure}

	\caption{
	Examples of traffic patterns from cloud IPs, identified by their UA strings. 
	The traffic from a user-initiated crawling bot meets our \massivelyshared criteria, whereas the others present constant or bursty traffic patterns, likely driven by different crawling targets at the CDN.}
	\label{fig:bot_traffic_examples}
\end{figure}

\PP{Showcasing Cloud Traffic} 
Before analyzing \massivelyshared traffic from cloud ASes systematically, we discuss common traffic patterns that we see from cloud IP addresses. 
As shown in Figure~\ref{fig:week_example} in Section~\ref{sec:method} and discussed in Section~\ref{sec:verification}, secure proxy services are often \massivelyshared and thus exhibit expected smooth and diurnal patterns similar to those in ISPs. 
In Figure~\ref{fig:bot_traffic_examples} we present two other examples: \textit{(a)} 
 Web crawlers of major search engines
 and \textit{(b)} 
 an interactive AI bot that 
accesses the Web on demand when users interact with the respective AI chatbot.
We see that the Web crawlers, Fig.~\ref{fig:not_shared_bot_traffic}, exhibit a variety of traffic patterns, none of which meet our criteria for \massivelyshared IPs. 
However, we find instances of bots that do, such as 
the on-demand AI crawler shown in Figure~\ref{fig:diurnal_search_engine_bot_traffic}. 
We will discuss these cases in more depth later in Section~\ref{sec:proxies_bots_cloud}.

\subsection{Toward Classifying \MassivelyShared Cloud IPs} 

As both proxy-like services and some instances of bots meet our criterion for \massivelyshared IPs, we dissect this set of cloud IPs with the goal of separating both cases. 

\PP{User-Agent Strings to Identify Users and Bots} 
We use the HTTP User-Agent (UA) strings in requests from \massivelyshared cloud IPs as an auxiliary data source to explore \textit{why} this behavior manifests in cloud traffic. Recall that the User-Agent field 
identifies the respective client application (e.g., a Web browser version, an app, a crawler or bot~\cite{rfc7231, mdn_browser_ua_detection, google_crawlers_user_agents, open_ai_crawlers}).

\PP{Homogeneous vs. Diverse IPs} 
Prior work found that IPs presenting few (or just a single) UA strings are more likely to be bots~\cite{richter_beyond_counting}, and a larger number of unique UA strings indicates end users and the various applications they use to access the Web~\cite{maier_handheld_device_traffic, kline_characteristics_ua, maier_nat_usage}. 
We build on this insight and tag IPs that issue requests with a single UA string as \textit{homogeneous}, \ie they come from bots, and IPs with multiple as \textit{diverse}, \ie they come from end-user proxies. 
We acknowledge that our separation is oversimplified, as bots may leverage multiple User-Agents, and cloud end-user gateways may be confined to single applications, leading proxy gateways to show up with a single UA. 
However, we present this as a first step toward teasing proxy-like IPs and bots apart.
\footnote{We refer the reader to Figure~\ref{fig:ua_string_hexbin_all_hosting} in  Appendix~\ref{app:ua_strings} for the full distribution of unique UA strings per IPv4 address and IPv6 /64s.}

Among \massivelyshared cloud IPv4 traffic, 81\% of \massivelyshared cloud traffic comes from \textit{diverse} IPs and 19\% comes from \textit{homogeneous} IPs. 
In IPv6, the contrast is even starker, with 96\% of \massivelyshared cloud traffic coming from diverse /64s and only 4\% from homogeneous /64s. 
We find similar proportions of IP addresses themselves, with 77\% of \massivelyshared IPv4 addresses and 84\% of \massivelyshared IPv6 /64s tagged as diverse.

\subsection{Of Proxies and Bots in the Cloud}
\label{sec:proxies_bots_cloud}

\PP{Diverse IPs -- Likely End-Users} 
First, analyze the cloud IPs that meet our \massivelyshared criteria and present multiple UA strings. 
Studying the ASes of \massivelyshared cloud IPs, we find that more than 70\% of all shared IPv4 traffic originates from 10 cloud providers, including hyperscalers, multi-tenants, CDNs, and cloud security companies. 
The top cloud AS (a hyperscaler) accounts for 17\% of the \textit{diverse} \massivelyshared IPv4 traffic from cloud providers. 
The second, a cloud security company, originates 15\% of this traffic and advertises secure web gateway services.

In IPv6, we see even higher levels of concentration, with 94\% of all \massivelyshared traffic coming from the top 10 cloud ASes.
 The top two providers with the most activity are large cloud providers that offer a multitude of services, including proxy services, Web hosting, and cybersecurity products. The top provider (a large CDN and cloud provider) is responsible for 37\% of all \massivelyshared cloud IPv6 traffic. 
Upon closer inspection, we find that all of the IPs in this provider are from a popular anonymization proxy service that publicly identifies its IPs through a geofeed. 

While not exhaustive, our analysis shows that a large amount of end-user \massivelyshared cloud traffic stems from proxy infrastructures.
 Importantly, these \massivelyshared deployments (in both IPv4 and IPv6) are likely also driven by anonymity, operational convenience, and business models centered on centralized traffic mediation and control.

\PP{Homogeneous IPs -- Likely Bots}
Next, we analyze the smaller fraction of \massivelyshared IPs from the cloud where UA strings are homogeneous, accounting for 19\% and 4\% of \massivelyshared IPv4 and IPv6 traffic, respectively. 
These cases are distinct from others of \massivelyshared IPs as these IPs meet the traffic shape criteria but are unlikely to directly connect end-users. 
Instead, we 
hypothesize that much of this activity reflects \textit{user-triggered} bot actions. 
That is, bots are triggered by link previews\cite{meta_crawlers, bing_crawlers, google_fetchers}, ad bidding~\cite{IMC2014-Back-Office}, user-triggered AI  crawlers~\cite{open_ai_crawlers, perplexity_crawlers, novaact}, or an application that initiates API calls~\cite{akamai_report}. 
While these types of IPs do not encompass end users as directly as proxies, they do capture a large amount of end-user browsing behavior that culminates in the same traffic shapes. 
Moreover, several companies that run these crawlers consider these to be user requests, thus not following the robots.txt protocol~\cite{open_ai_crawlers, perplexity_crawlers, robotx.txt_rfc9309}. 

Indeed, in IPv4, the top cloud AS originating homogeneous, \massivelyshared traffic (28\% of traffic), the most common UA is from a popular business application suite that interacts with web services for the user; the second 
 is associated with a large e-commerce platform that fetches Web resources. In the second most popular AS (14\% of traffic), the only UA string presented is for an advertising platform that is likely collecting analytics for their customers or carries out ad bidding.

\begin{tcolorbox}[mytakeaway]
	\textbf{Takeaway:} A substantial amount of cloud traffic from \massivelyshared IP addresses is end-user proxies, including consumer VPN/exit services and enterprise secure gateway deployments. A smaller amount of this traffic comes from bots, many of which appear to be user-triggered.
\end{tcolorbox}

%% file: results_over_time.tex
\section{\MassivelyShared IPv4 Over Time}
\label{sec:sharing_over_time}

Finally, we study how \massive address sharing has evolved in IPv4. 
For each month, we chose a week that does not include a major holiday globally and repeat our methodology. We observe that, \textit{over the course of two years, the fraction of overall IPv4 traffic from massively shared IPs increases by nearly 10\%}, ranging from 36\% in January 2024 to 45\% in October 2025.

Specifically in ISPs, 
\massivesharing in IPv4 increases in all regions, with Asia, Africa, and South America starting at higher rates of \massivelyshared traffic out of all traffic (45\%, 43\%, 45\%, respectively) and increasing in traffic from \massivelyshared IPv4 addresses (in Asia 11\%, in Africa 8.8\%, and in South America 7.7\%). 
We also observe \massivesharing in North America, Oceania, and Europe, starting at 24\%, 16\%, and 26\% of all traffic and increasing by 10\%, 4.5\%, and 0.9\%, respectively. 
Given that Internet penetration is increasing~\cite{itu_ict_statistics}, we can expect this trend to continue.

\begin{tcolorbox}[mytakeaway]
	\textbf{Takeaway:} The fraction of \massivelyshared IPv4 traffic out of all IPv4 traffic has increased by 10\% between early 2024 and late 2025. While pronounced in regions with scarce IPv4 resources, we see increasing \massivesharing globally.
\end{tcolorbox}

%% file: conclusion.tex
\section{Concluding Discussion}
\label{sec:conclusion}
In this work, we characterized large-scale IP sharing, which we coined \massivelyshared IPs. 
We analyzed the prevalence of this deployment over both IPv4 and IPv6 traffic as seen globally by a major CDN. 
Our method for \massivesharing detection relies solely on traffic shape, detectable by Fourier analysis and curve-fitting evaluation.  
We find that over 40\% of IPv4 requests to a major CDN come from \massivelyshared IP addresses, deployed across IPv4 networks globally, and increasingly over time. 
In this section, we discuss some of the key takeaways from our work. 

\PP{IPv4 Sharing and IPv6 Deployment} 
\Massivelyshared IPv4 is widely deployed in End-User ISPs (Section~\ref{sec:sharing_overview}), and about half of this traffic originates from addresses that also have IPv6 connectivity (Section~\ref{sec:isps:ipv6_capabilities}). 
This suggests 
these networks are prepared to reduce reliance on \massivelyshared IPv4. 
However, IPv4-only deployments and partial IPv6 deployments are prevalent, particularly in Africa, introducing another single point of failure between clients and servers. 
Unless IPv6 deployment accelerates, the Internet will continue to broadly rely on IPv4 sharing, exacerbating reliance on a relatively small quantity of Internet resources.

\PP{\MassiveSharing Is Not Always Driven By Scarcity} 
Contrary to initial intuition, we find many cases of \massivelyshared addressing that are \textit{not} driven by IPv4 scarcity. In fact, we observe \massivelyshared IPv6 /64s in major ISPs (Section~\ref{sec:isps:ipv6}), as well as many instances of \massivelyshared IPv4 (and to some degree IPv6) in the cloud (Section~\ref{sec:cloud}), underpinning proxy, enterprise gateway, and anonymization services. 
From this, we point to broader shifts in Internet infrastructure, illustrating that large quantities of end-user traffic originate outside of ISPs. 

\PP{Massive IP Sharing Is Here To Stay} 
From both of these observations, one thing becomes clear: IPv4 exhaustion and IPv6 deployment aside, \massivesharing is widespread, increasing (Section~\ref{sec:sharing_over_time}), and if observed trends continue, we may see even higher proportions of traffic carried over \massivelyshared IP addresses in the future. \Massivelyshared addressing complicates security, geolocation, host reputation, and blocklisting, underscoring the need for robust sharing-detection techniques and for security systems to move beyond IP-based attribution. 
To combat these, detection methods and characterization must be continuously deployed and studied.

\PP{Future Work}
\label{sec:future_work}
As our method is purely centered on traffic shape, 
future work can apply it to other vantage points, including transit networks and sampled telemetry such as IPFIX, NetFlow, and sFlow, and others  
in large-scale network management and those with limited access to other traffic details. 
 Future work may fine-tune parameters for specific traffic types and vantage points, and exploit additional FFT properties such as frequency magnitudes and amplitudes or incorporate weekday–weekend variations.

%% file: acknowledgments.tex
\section{Acknowledgements}
The authors are grateful for the assistance and support of KC Ng. 
This material is based upon work supported by the National Science Foundation (NSF) Graduate
Research Fellowship Program under Grant No. 2545526 and NSF CNS award 2319315. Any opinions,
findings, and conclusions or recommendations expressed in this material are those of the
author(s) and do not necessarily reflect the views of the National Science Foundation.

%% file: appendix.tex
\clearpage

\section{Ethics}
\label{app:ethics}
Measurements in this study were conducted using confidential operational data provided by the participating CDN. Data processing took place exclusively on secure machines maintained and controlled by the CDN. Access to the data was restricted to authorized personnel with valid credentials, and all activities adhered to the CDN's internal policies, security controls, and applicable regulatory requirements.  The reported results are aggregated such that no personal identifying information of any end-user could be inferred, nor any proprietary information of the CDN, such as individual customers or network partners.

\section{Parameter Sensitivity Analysis}
\label{sec:sensitivity}
\label{app:method_details}

\begin{figure}
	\includegraphics[width=.49\linewidth]{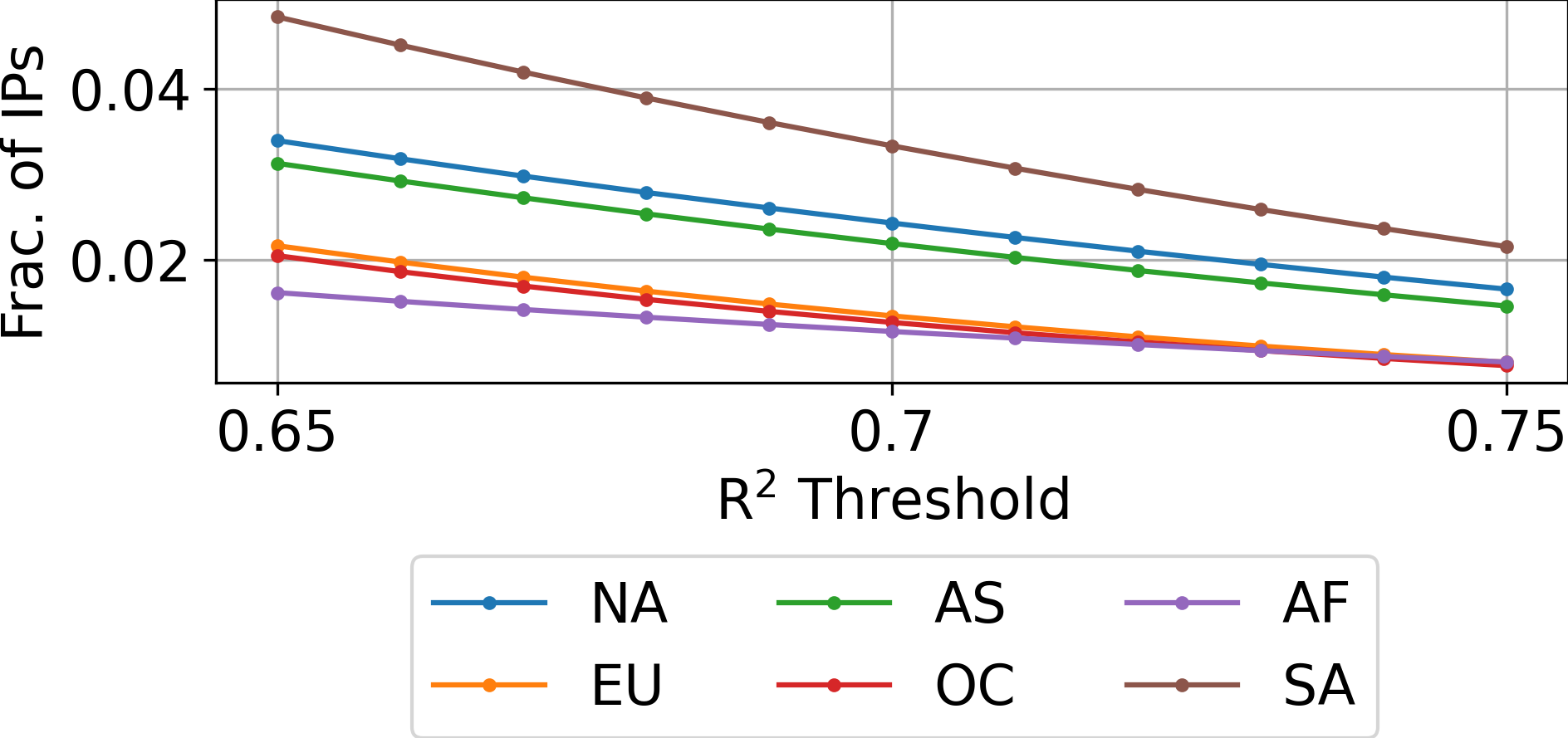}
	\caption{The fraction of active IPs that are \massivelyshared with different $R^{2}$ thresholds per continent. We experiment with values from 0.65 to 0.75 in increments of 0.01.}
	\label{fig:r_squ_sensitivity_by_continent}
\end{figure}

$\boldsymbol{R^{2}}$ \textbf{ Sensitivity.}
\label{sec:r_squ_sensitivity}
Next, we experiment with how the $R^{2}$ threshold affects the fraction of \massivelyshared IPs and traffic we detect. 
In Figure~\ref{fig:r_squ_sensitivity_by_continent}, we present how the fraction of \massivelyshared IPs changes per continent with different thresholds, seeing that this fraction does change. 
 This presents a trade-off between the strictness of a definition of smoothness (\ie a higher $R^{2}$ value is stricter) and the number of IPs we capture. 
We note that this pattern also applies to the fraction of traffic from these IPs to a higher degree. When varying the value from 0.65 to 0.75, the total \massivelyshared traffic volume changes by 7-12\% (depending on continent). 
We proceed with using 0.7 as our threshold based on our experimentation, but emphasize that in future work this can be adjusted based on the nature of the traffic. 

\PP{Stability over Time}
The bulk of our analysis is performed over data from a 3-day period in a single non-holiday week in July 2025. 
While we analyze a relatively small time period, we observe that our results are consistent when repeated over the course of 1 month (throughout 4 weeks in May 2025). 
When we experiment per continent, the fraction of \massivelyshared IPs changes by at most 0.2\% over 1 month, and the fraction of \massivelyshared traffic changes by at most 3\%. 
Therefore, our chosen time period does not substantially affect our results. 

\begin{figure}
        \includegraphics[width=.49\linewidth]{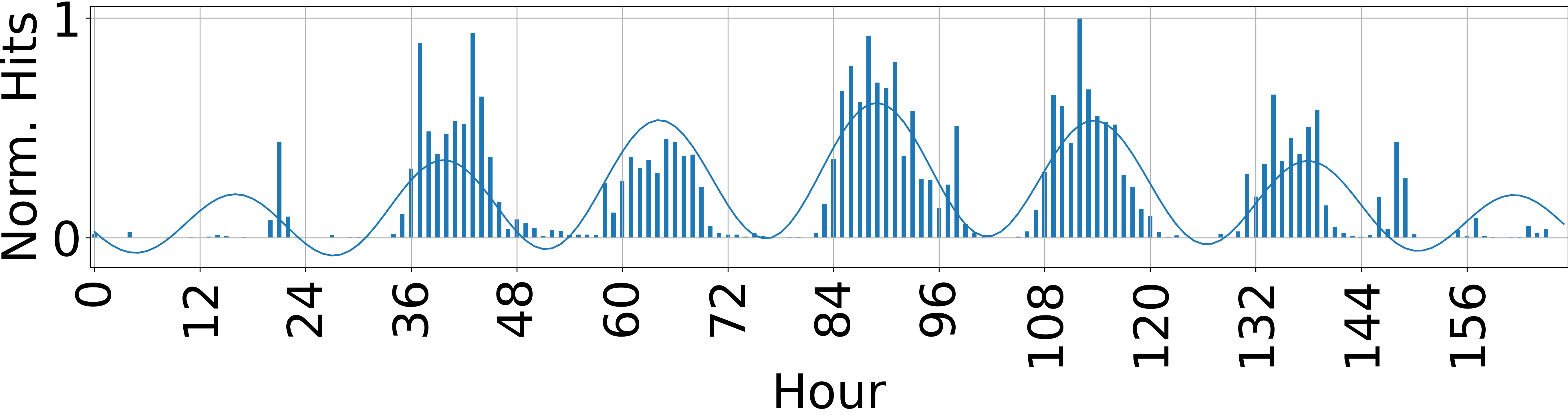}
	\caption{7 days of traffic from an IP in an enterprise proxy service. We see little activity over the weekend, motivating our choice to select Tuesday-Thursday as the 72-hour period we analyze weekly.}
	\label{fig:week_example}
\end{figure}

\section{Implementation: Time Period}
\label{app:time_period}
Figure~\ref{fig:week_example} shows a case of this disruption in an enterprise gateway. 
The full 168 hours of traffic does not meet our criteria due to poor fit. However, isolating to Tuesday-Thursday (72 hours) does. 
We find several such examples in our exploration and therefore apply our method to 72 hours of consecutive traffic from 
Tuesday to Thursday. 

\begin{figure}
	\centering
	\begin{subfigure}[b]{.49\linewidth}
		\includegraphics[width=\linewidth]{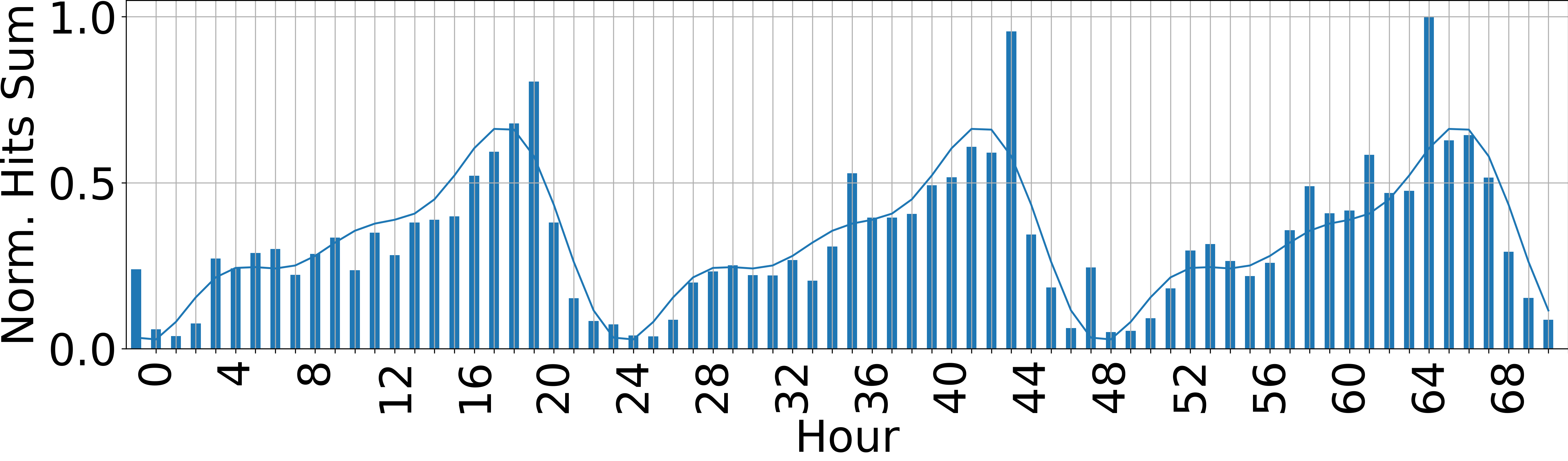}
		\caption{ISP in Germany ($R^{2} = 0.77$)}
		\label{fig:de_agg_ex}
	\end{subfigure}
	\begin{subfigure}[b]{.49\linewidth}
		\includegraphics[width=\linewidth]{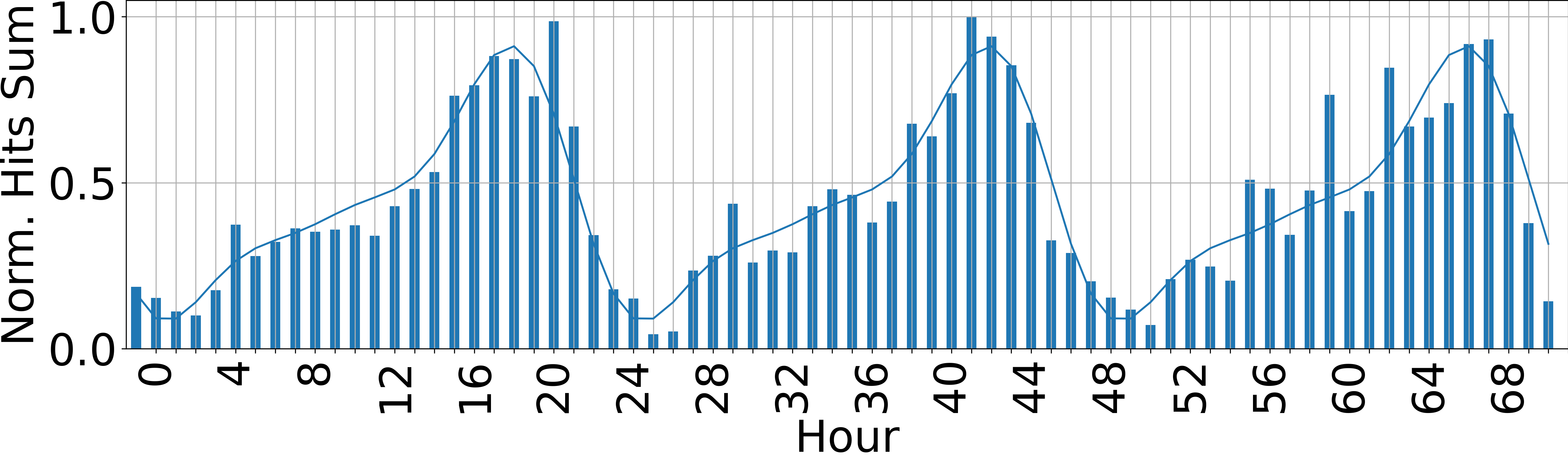}
		\caption{ISP in France ($R^{2} = 0.87$)}
		\label{fig:fr_agg_ex}
	\end{subfigure}
	\begin{subfigure}[b]{.49\linewidth}
		\includegraphics[width=\linewidth]{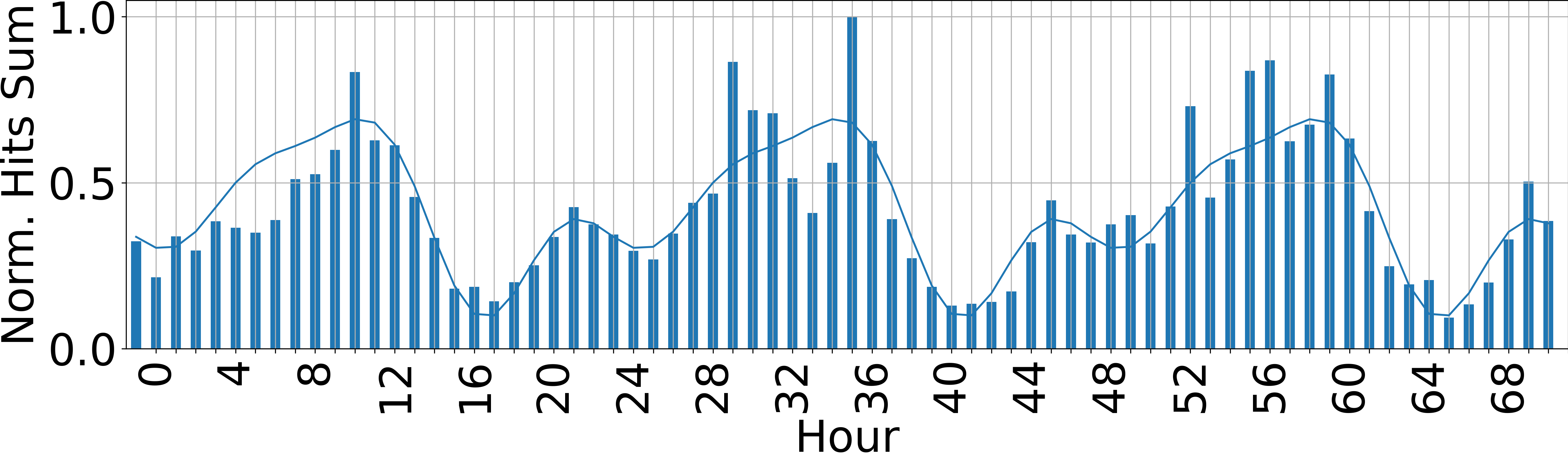}
		\caption{ISP in Japan ($R^{2} = 0.75$)}
		\label{fig:jp_agg_ex}
	\end{subfigure}
	\begin{subfigure}[b]{.49\linewidth}
		\includegraphics[width=\linewidth]{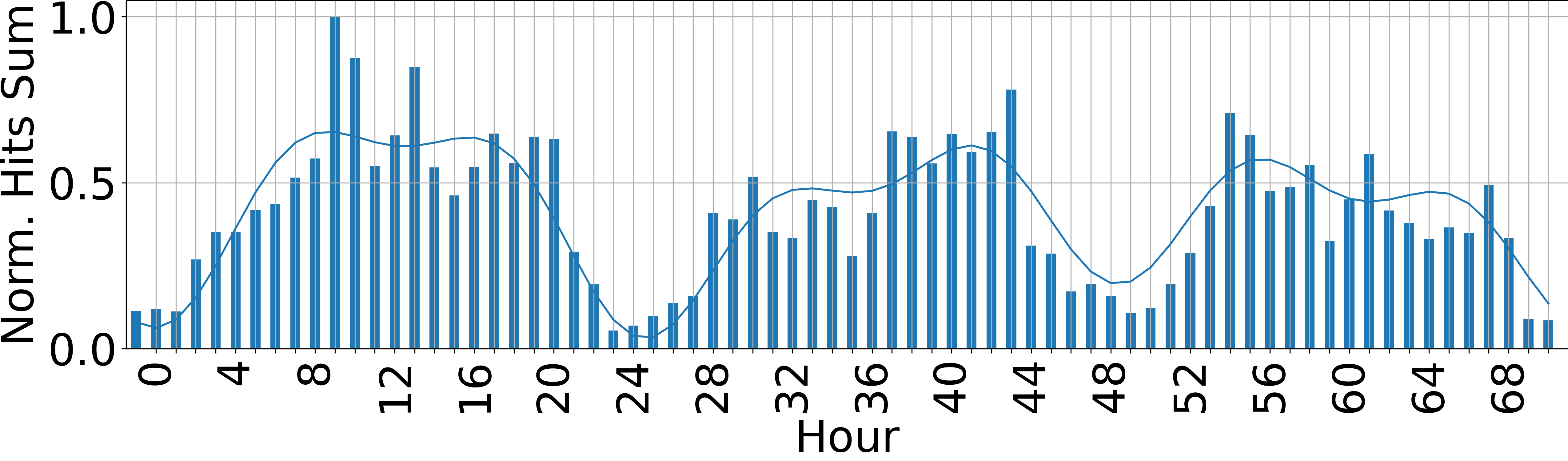}
		\caption{ISP in Iran ($R^{2} = 0.70$)}
		\label{fig:ir_agg_ex}
	\end{subfigure}
	\caption{We present traffic aggregated from /24s in 4 different countries with varying levels of CDN presence. We find that all traffic aggregated together meets our \massivelyshared criteria.}
	\label{fig:agg_exp_countries}
\end{figure}

\begin{figure}
	\centering
	\begin{subfigure}[b]{.49\linewidth}
		\includegraphics[width=\linewidth]{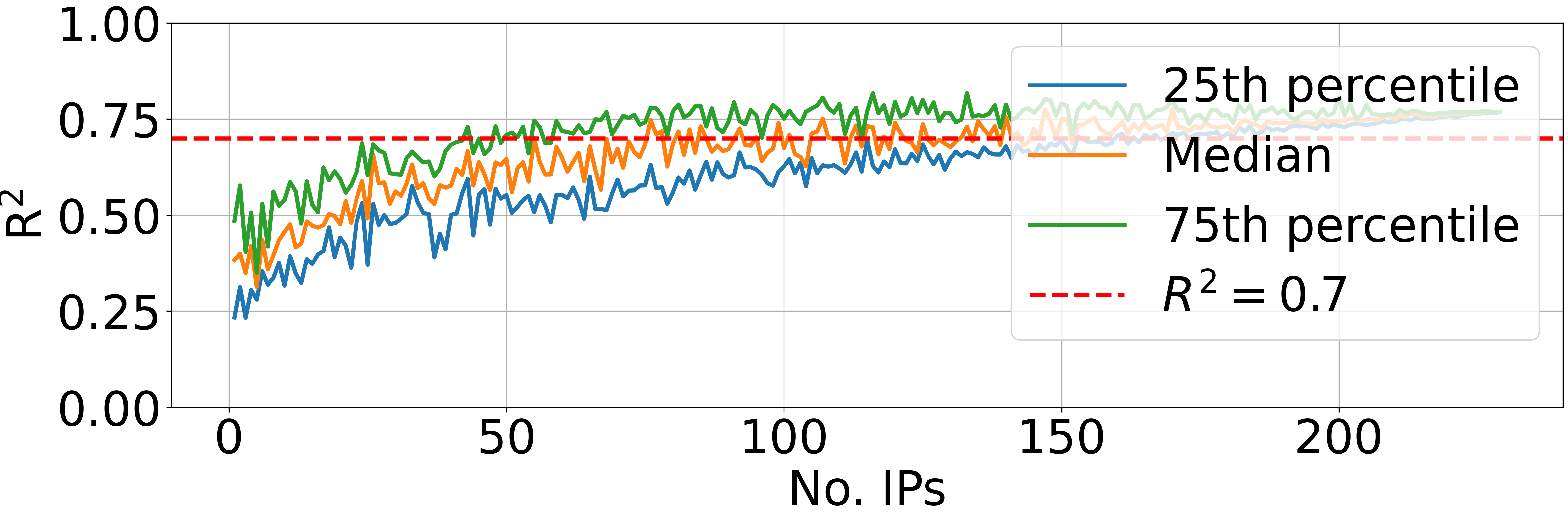}
		\caption{ISP in Germany}
		\label{fig:de_isp_r_squ_exp_lines}
	\end{subfigure}
	\begin{subfigure}[b]{.49\linewidth}
		\includegraphics[width=\linewidth]{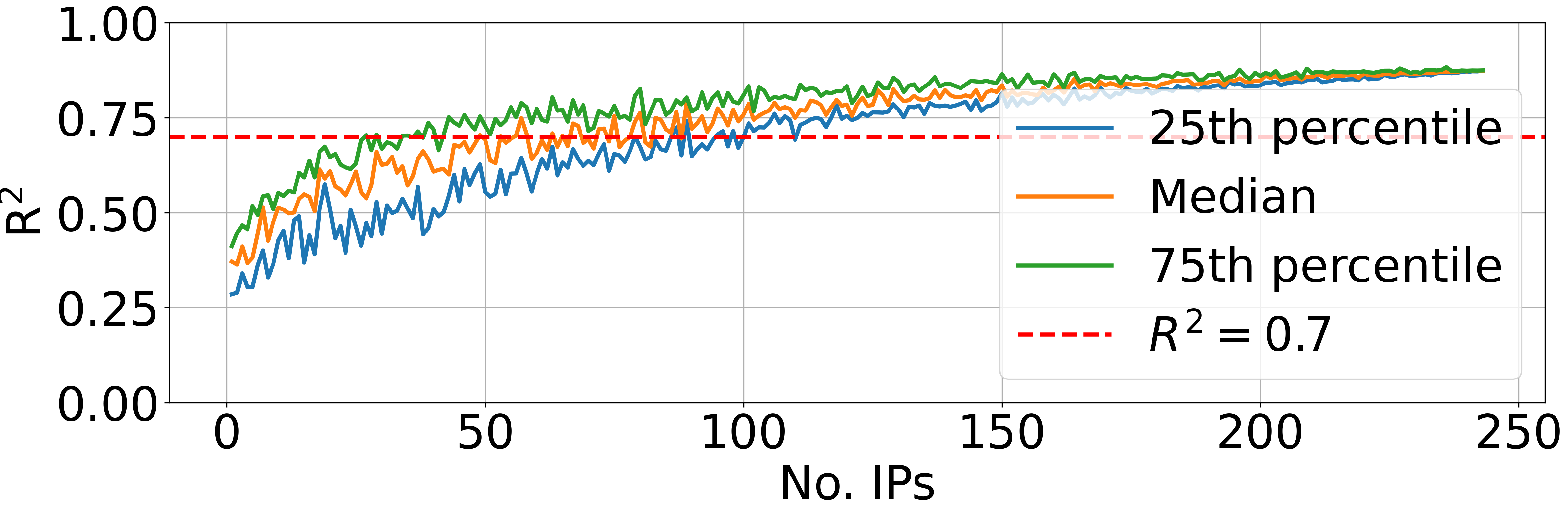}
		\caption{ISP in France}
		\label{fig:fr_isp_r_squ_exp_lines}
	\end{subfigure}
	\begin{subfigure}[b]{.49\linewidth}
		\includegraphics[width=\linewidth]{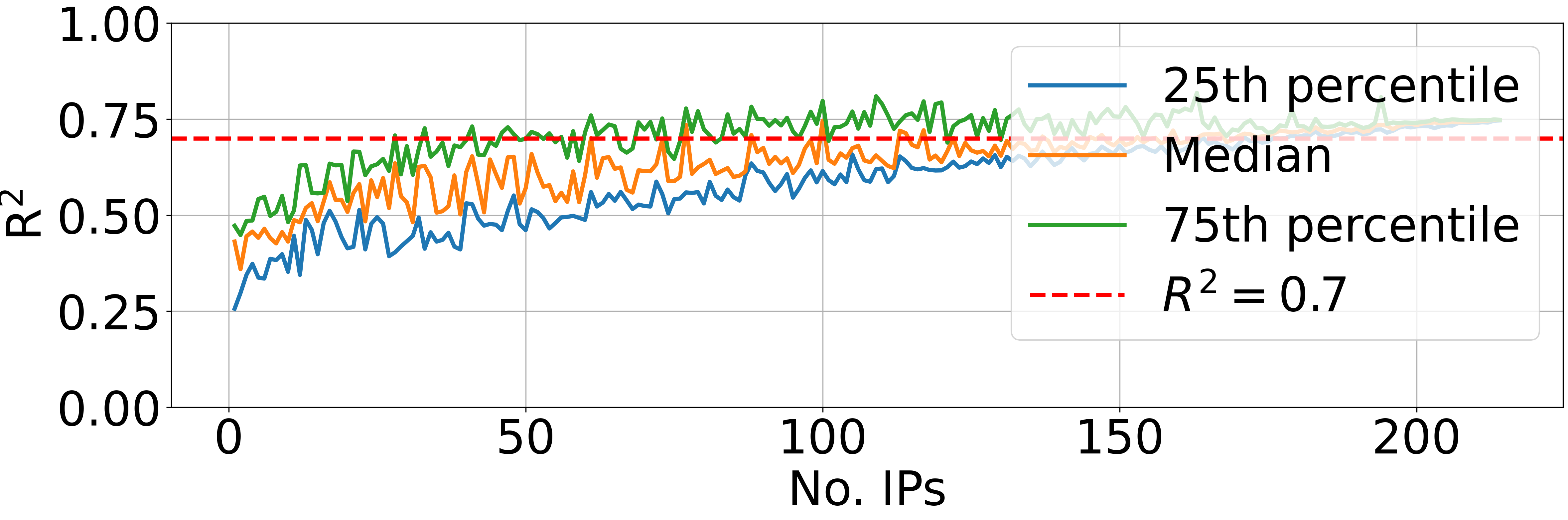}
		\caption{ISP in Japan}
		\label{fig:jp_isp_r_squ_exp_lines}
	\end{subfigure}
	\begin{subfigure}[b]{.49\linewidth}
		\includegraphics[width=\linewidth]{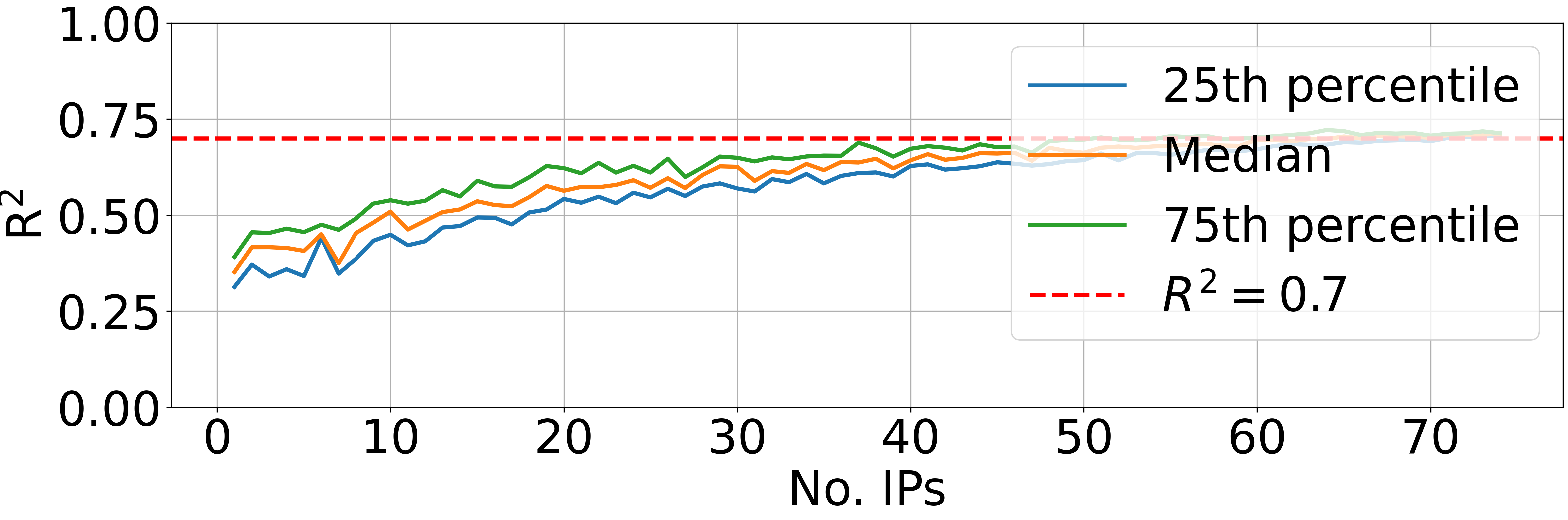}
		\caption{ISP in Iran}
		\label{fig:ir_isp_r_squ_exp_lines}
	\end{subfigure}
	\caption{We repeat the aggregation experiment presented in Figure~\ref{fig:r_squ_increase_example} in /24s in 4 different countries. We find that all traffic aggregated together meets our \massivelyshared criteria between 65-75 individual IPs.}
	\label{fig:subscriber_estimate_agg_exp_countries}
\end{figure}

\section{Examples From Other Countries}
\label{app:other_countries}
We also perform the multi-user simulation experiment in Section~\ref{sec:motivating_ex} in /24s from networks in other countries and find similar results, that aggregating individual IPs from each network eventually reaches our \massivelyshared criteria. 
We identify these IPs through research of the networks and through networks identified through RIPE Atlas~\cite{ripe_atlas_platform}. 
We identify probes that are tagged with ``Native IPv4.'' 
Although we do not have ground truth that these IPs in fact do have native IPv4 or that these ISPs use single-subscriber IPv4 addressing, we highlight that this aligns with our claims of our estimations being lower bounds. 
That is, if more subscribers are behind individual IPs, this means that more user traffic was necessary to reach our criteria. 

We show the results in Figure~\ref{fig:agg_exp_countries}. 
We observe that this experiment is successful in multiple distinct regions globally. 
We specifically highlight the network in Iran, where we see significantly fewer IPs total active in the /24 than in the networks of other countries. 
We hypothesize that one reason for this lack of activity from IPs is due to both the CDN's presence in the region and potential Internet disruptions~\cite{cloudflareIranInternetPartiallyRestored2026}. 
Despite this, we are still able to aggregate traffic that meets our diurnal and smooth criteria. 
This shows that when the traffic of many users is aggregated, these signals can appear even with sparse data. 
We also use this to build confidence in other geographic areas that appear similarly in the CDN's access logs. 

\PP{Estimating Single Subscriber Count Behind \MassivelyShared IPs}
To further quantify how many single subscribers are captured in \massivelyshared IPs, we repeat the analysis presented in Section~\ref{sec:quantifying_usage}. 
We show the results of these experiments in Figure~\ref{fig:subscriber_estimate_agg_exp_countries}. 
In these examples specifically, 
we find that the median number of single subscriber IPs we need to aggregate to meet our \massivelyshared criterion is 65-75 in all cases. 
We remind the reader that this is the lower bound for the number of subscribers needed to meet this criterion; adding more single subscriber IPs past this amount consistently meets our criteria. 

Of course, this method has caveats, as we have no ground truth for these particular networks, only context from operator reports on RIPE Atlas and ISP documentation. 
However, as stated in Section~\ref{sec:quantifying_usage}, we provide this as a heuristic to estimate subscriber count behind \massivelyshared IPs.  

\section{Manual Validation of Known Technologies}
\label{app:small_validation}
In addition to the widespread validation across large proxy geofeeds or entire countries described in Section~\ref{sec:verification}, we additionally validate individual IPs by manually collecting IPs. 
We collect cellular IPs from 4 different providers across the US and Europe, and IPs from 2 different proxy services from our colleagues. 
In all cases, we find that these cases meet our \massivelyshared criteria. 
Although small in scale, these cases provide ground truth validation. 

\section{UA String diversity}
\label{app:ua_strings}
\begin{figure}
	\begin{subfigure}[]{0.45\linewidth}
		\includegraphics[width=\linewidth]{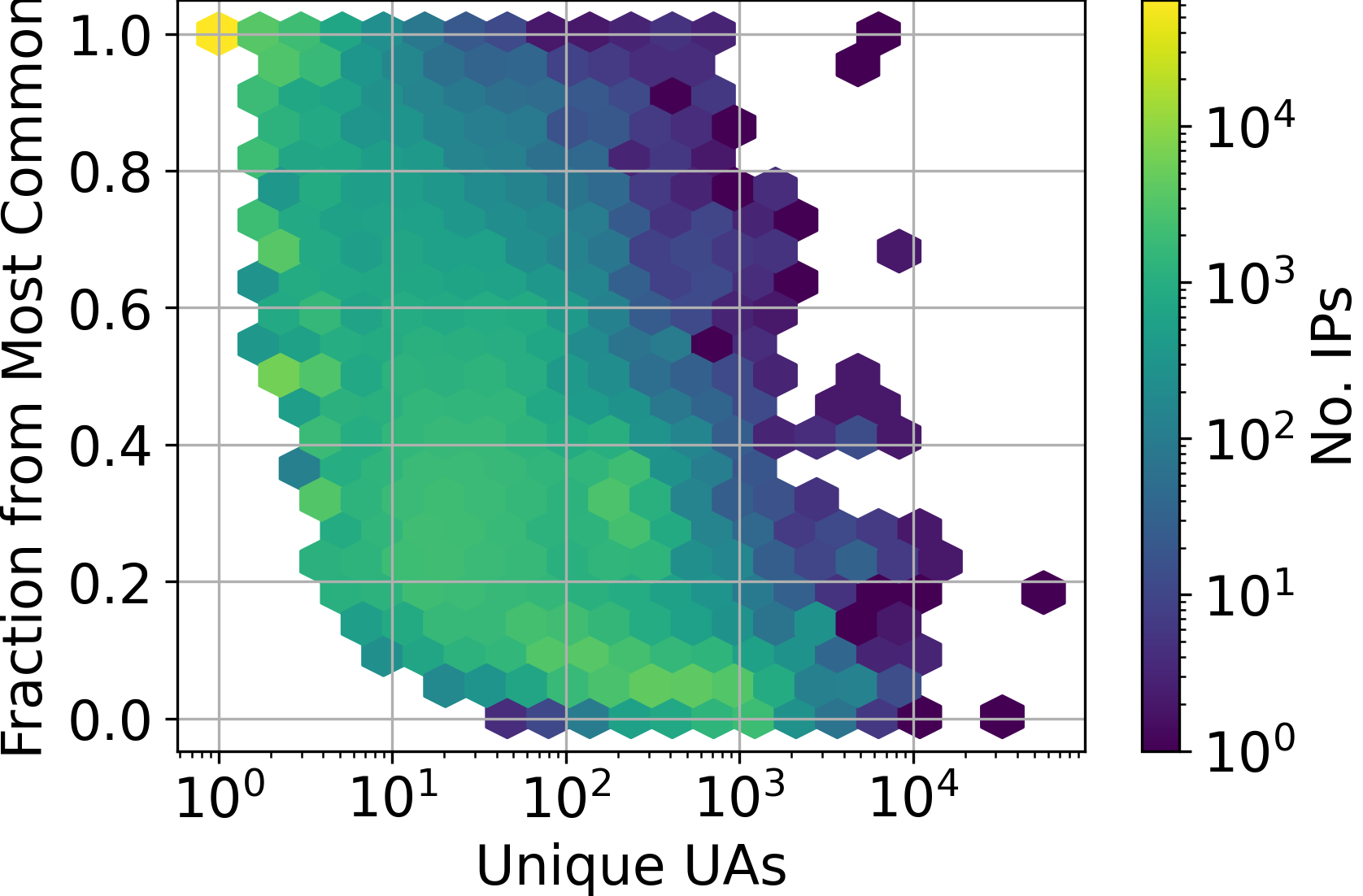}
		\caption{IPv4}
		\label{fig:ipv4_ua_string_hexbin_all_hosting}
	\end{subfigure}
	\begin{subfigure}[]{0.45\linewidth}
		\includegraphics[width=\linewidth]{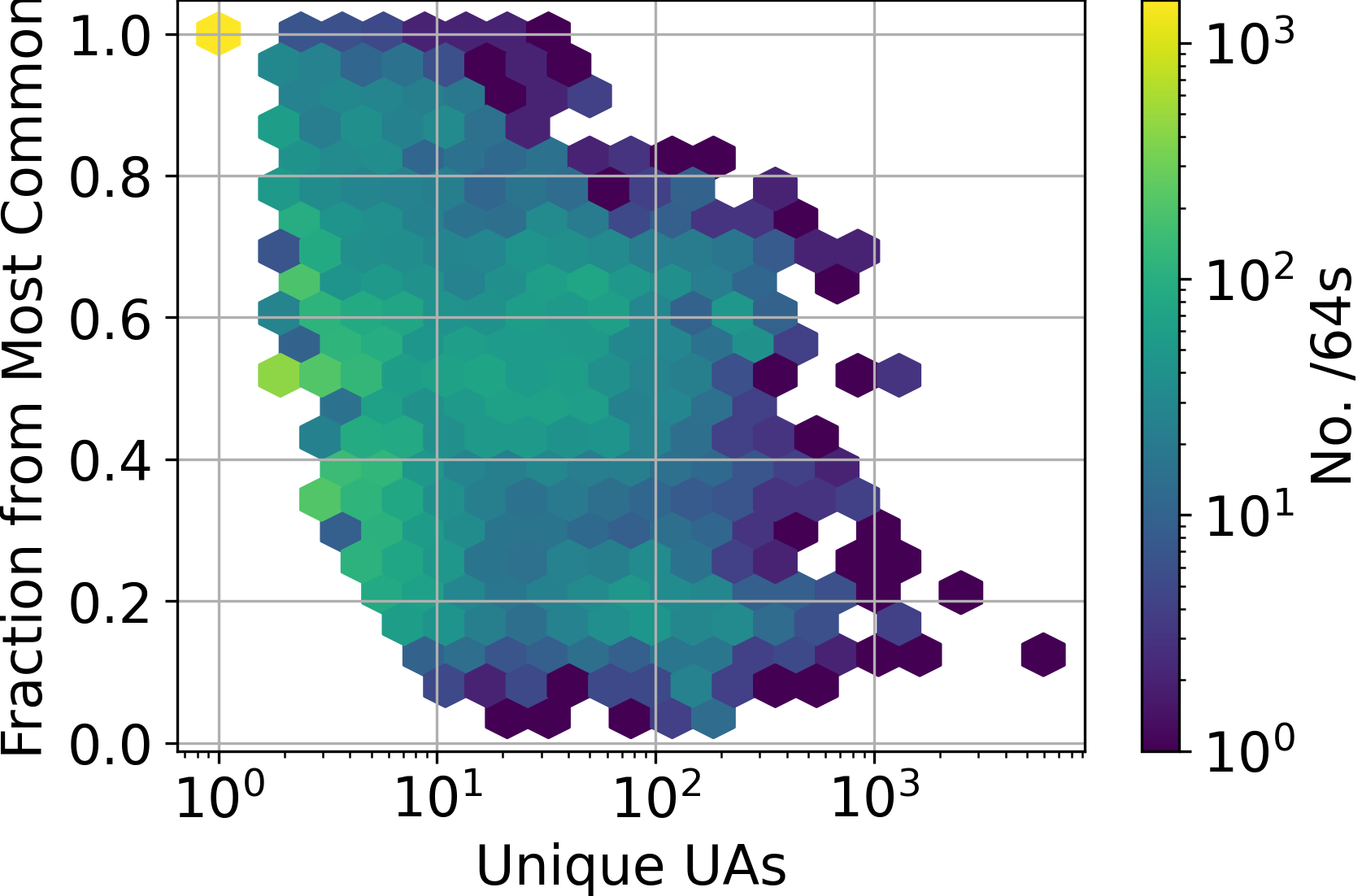}
		\caption{IPv6}

		\label{fig:ipv6_ua_string_hexbin_all_hosting}
	\end{subfigure}
	\caption{UA string diversity from \massivelyshared traffic in cloud providers.
	On the y-axis, we plot the fraction of UA strings that the most common UA string represents. For example, if we observe 10 hits from an IP, and 9 of them have the same UA string, the fraction will be 0.9. On the x-axis, we compare this to the total unique UA strings we observe for this IP (note the log scale).
	We observe a concentration of IPs that present a single UA string, which prior work has identified as a bot characteristic~\cite{richter_beyond_counting}.}
	\label{fig:ua_string_hexbin_all_hosting}
\end{figure}

Here, we present a more detailed view of UA string diversity across \massivelyshared IPs in cloud providers. 
In Figure~\ref{fig:ua_string_hexbin_all_hosting}, we present the UA strings presented by these IPs and /64s. 
We observe that a large fraction of IPs (19.2\% in IPv4, 4.2\% in IPv6) present a single UA string, which we refer to in Section~\ref{sec:cloud} as \textit{homogeneous} IPs. 

